\documentclass[
pre,
aps,
twocolumn,
english,
superscriptaddress,
floatfix,
amsmath,amssymb,
longbibliography,
nofootinbib
]{revtex4-2}

\usepackage{graphicx}
\usepackage{dcolumn}
\usepackage{bm}
\usepackage[hidelinks]{hyperref}
\usepackage{soul}
\hypersetup{colorlinks = true, allcolors = blue}

\usepackage{xcolor}

\graphicspath{{./Figures/}}

\begin{document}

\preprint{APS/123-QED}

\title{Capacitive response of biological membranes}

\author{Jafar Farhadi}
\thanks{These authors contributed equally.}
\affiliation{Department of Chemical and Biomolecular Engineering, University of California, Berkeley, CA 94720, USA}

\author{Joshua B. Fernandes}
\thanks{These authors contributed equally.}
\affiliation{Department of Chemical and Biomolecular Engineering, University of California, Berkeley, CA 94720, USA}
\affiliation{Chemical Sciences Division, Lawrence Berkeley National Laboratory, CA 94720, USA}

\author{Karthik Shekhar}
\email{kshekhar@berkeley.edu}
\affiliation{Department of Chemical and Biomolecular Engineering, University of California, Berkeley, CA 94720, USA}
\affiliation{Helen Wills Neuroscience Institute, California Institute for Quantitative Biosciences, QB3, Center for Computational Biology, University of California, Berkeley, CA 94720, USA}
\affiliation{Biological Systems Division, Lawrence Berkeley National Laboratory, Berkeley, CA 94720, USA}

\author{Kranthi K. Mandadapu}
\email{kranthi@berkeley.edu}
\affiliation{Department of Chemical and Biomolecular Engineering, University of California, Berkeley, CA 94720, USA}
\affiliation{Chemical Sciences Division, Lawrence Berkeley National Laboratory, CA 94720, USA}

\date{\today}

\begin{abstract}
\vspace{0.1in}
We present a minimal model to analyze the capacitive response of a biological membrane subjected to a step voltage via blocking electrodes. Through a perturbative analysis of the underlying electrolyte transport equations, we show that the leading-order relaxation of the transmembrane potential is governed by a capacitive timescale, 
${\tau_{\rm C}
=\dfrac{\lambda_{\rm D}L}{D}\left(\dfrac{2+\Gamma\delta^{\rm M}/L}{4+\Gamma\delta^{\rm M}/\lambda_{\rm D}}\right)}$, where $\lambda_{\rm D}$ is the Debye screening length, $L$ is the electrolyte width, $\Gamma$ is the ratio of the dielectric permittivity of the electrolyte to the membrane, $\delta^{\rm M}$ is the membrane thickness, and $D$ is the ionic diffusivity. 
This timescale is considerably shorter than the traditional RC timescale 
${\lambda_{\rm D} L / D}$ 
for a bare electrolyte due to the membrane's low dielectric permittivity
and finite thickness. 
Beyond the linear regime, however, salt diffusion in the bulk electrolyte drives a secondary, nonlinear relaxation process of the transmembrane potential over a longer timescale ${\tau_{\rm L} =L^2/4\pi^2 D}$.
A simple equivalent-circuit model accurately captures the linear behavior, and the perturbation expansion remains applicable across the entire range of observed physiological transmembrane potentials. 
Together, these findings underscore the importance of the faster capacitive timescale and nonlinear effects on the bulk diffusion timescale in determining transmembrane potential dynamics for a range of biological systems. 
\end{abstract}

\maketitle

\section{Introduction}
\label{sec:introduction}

Biological membranes are ubiquitous in living systems, where they separate electrolyte compartments with different ionic compositions. In physiological conditions, when exposed to electric fields, biological membranes can develop transmembrane potentials ranging from tens of millivolts to a few volts \cite{hodgkin1952quantitative,hille1992,tsong1991electroporation,kotnik2019membrane}. Alterations in the transmembrane potential involve the dynamic charging and discharging of the membrane, which is governed by the diffuse charge layers at the membrane-electrolyte interface \cite{row2025spatiotemporal}. Because these processes involve the transport of ions in both the bulk electrolyte and the diffuse layers, a central question arises regarding the capacitive behavior of biological membranes: \textit{What are the characteristic timescales associated with their charging and discharging?}

\begin{figure}
    \centering
    \includegraphics[width=\linewidth]{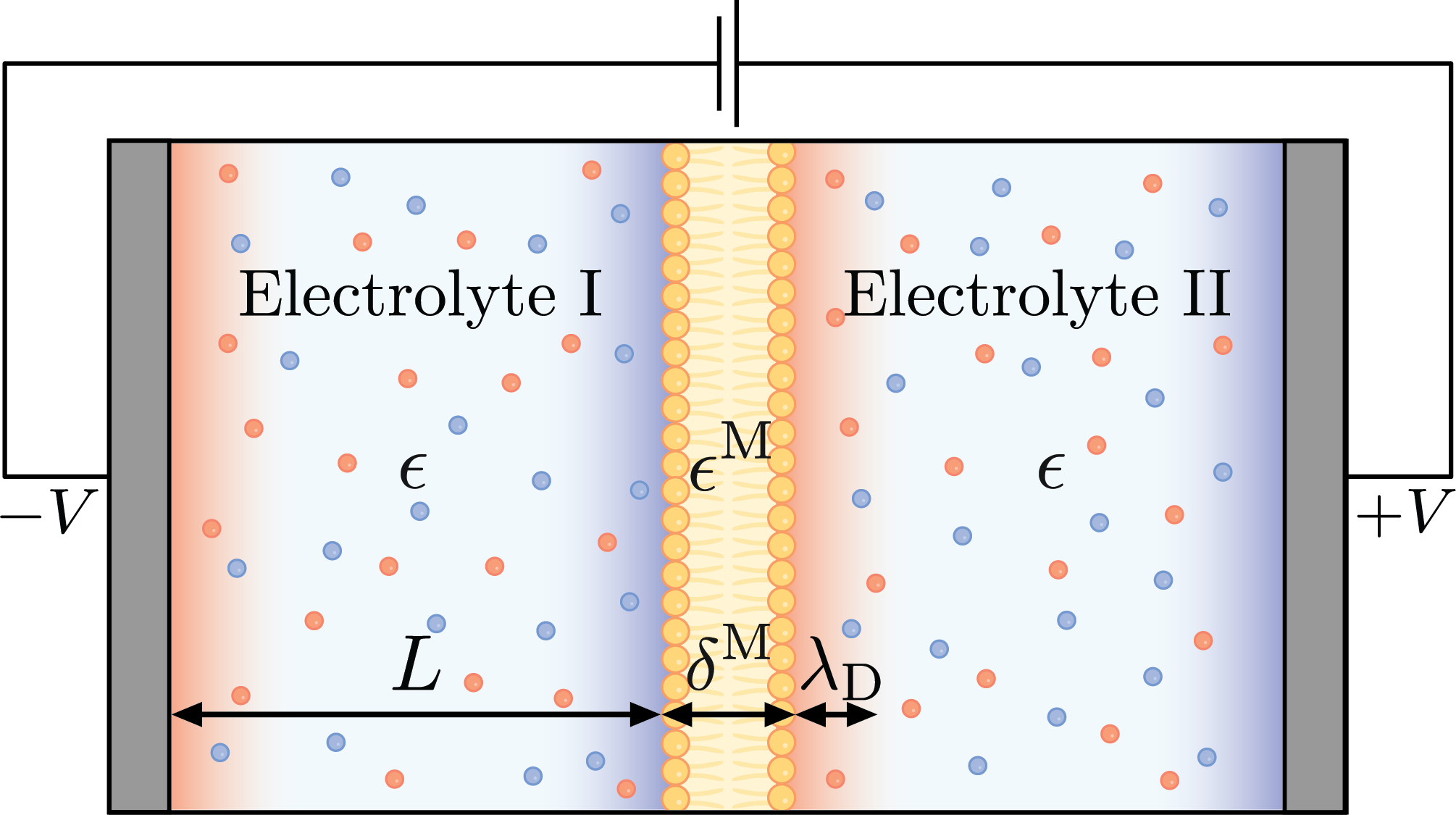}
    \caption{Model schematic. Two domains (I and II) of size $L$, each containing a symmetric binary electrolyte solution of permittivity $\epsilon$, are separated by an impermeable membrane of thickness $\delta^{\rm M}$ and permittivity $\epsilon^{\rm M}$. A voltage difference of $2V$ is applied via two blocking electrodes. Consequently, four diffuse charge layers develop (one at each interface), each with characteristic thickness equal to the Debye length $\lambda_{\rm D}$.}
    \label{fig:fig_schematic}
\end{figure}

In this work, we study the capacitive response of an impermeable planar membrane with finite thickness, $\delta^\mathrm{M}$, separating two symmetric binary electrolyte solutions of thickness $L$; see Fig.~\ref{fig:fig_schematic}.
We apply a step voltage using two blocking electrodes and examine the timescales governing the system's relaxation to a steady state.
Our main result is that following the rapid formation of diffuse charge layers on a short timescale ${\sim \lambda_{\rm D}^2 / D}$, the leading-order relaxation behavior of the transmembrane potential occurs over the capacitive timescale
\begin{align}
    \tau_{\rm C} &= \frac{\lambda_{\rm D}L}{D} \frac{\left(2+\Gamma\delta^{\rm M}/L\right)}{\left(4 + \Gamma\delta^{\rm M}/\lambda_{\rm D}\right)} \ , \label{eq:dim-tauC}
\end{align}
where $\lambda_{\rm D}$ is the Debye screening length, $D$ is the diffusion constant, and ${\Gamma=\epsilon/\epsilon^{\rm M}}$ is the dielectric mismatch between the electrolyte and the membrane. Following the initial linear relaxation over $\tau_{\rm C}$, we find that nonlinear effects become significant, manifesting as a secondary relaxation process controlled by the slower diffusion of salt in the bulk electrolyte on a timescale ${\sim L^2/D}$. Consequently, the full relaxation process depends nonlinearly on the applied potential and the dielectric properties of the membrane.

Our work builds on the foundational model problem considered by Bazant, Thornton, and Ajdari \cite{Bazant04}, who analyzed the transient response of a \textit{single} binary electrolyte subjected to an applied voltage via blocking electrodes. They showed that the overall relaxation to equilibrium is governed by the RC timescale of the bare electrolyte ${\tau_{\rm B}\equiv \lambda_{\rm D}L/D}$, a result that can be traced back to earlier works \cite{macdonald1974binary,kornyshev1981conductivity}. From Eq.~\eqref{eq:dim-tauC}, it is readily seen that when the membrane is absent (${\delta^{\rm M}\to 0}$), the capacitive timescale approaches the bare electrolyte RC timescale (${\tau_{\rm C}\sim \tau_{\rm B}}$). However, in the presence of a low-dielectric membrane with finite thickness, one finds that ${\tau_{\rm C}\ll\tau_{\rm B}}$. Thus, the presence of a membrane substantially speeds up the charging/discharging process.

This paper is organized as follows.
In section~\ref{sec:problem_statement}, we describe the model and non-dimensionalize the governing equations and boundary conditions.
In section~\ref{sec:numerics}, we present numerical observations demonstrating both the linear and nonlinear regimes, as well as their associated relaxation timescales.
In section \ref{sec:results}, we first outline a perturbation approach to obtain analytical solutions. We then present the linear dynamics to identify the membrane-mediated capacitive timescale (Eq.~\eqref{eq:dim-tauC}) and construct an equivalent circuit that is valid even for voltages larger than the thermal voltage (${\sim 25 \text{ mV}}$).
Next, we analyze the nonlinear behavior driven by the diffusion of ions in the electrolyte, which governs long-time relaxation. Finally, we compare the equilibrium steady-state achieved at long times with the classical solution obtained by Gouy \cite{Gouy1910} and Chapman \cite{chapman1913li}.
In section \ref{sec:discussion}, we discuss the relevance of these results to experimental settings and highlight the significance of the capacitive timescale. A connection with our recent work \cite{row2025spatiotemporal} suggests that $\tau_{\rm C}$ may play a fundamental role in the electrochemical relaxation of localized ion channel currents, which underlie the dynamic regulation of transmembrane potential in biological cells.

\section{Problem Statement}
\label{sec:problem_statement}

Consider a planar, uncharged lipid membrane (M) separating two domains each containing a symmetric dilute electrolyte solution confined between two blocking electrodes, at a distance of ${2L + \delta^\mathrm{M}}$ from each other (see Fig.~\ref{fig:fig_schematic}). The electrolyte concentrations are considered to be ${\sim 150\text{ mM}}$, typically found in many physiological conditions \cite{lodish2000molecular}. 
The membrane thickness, ${\delta^\mathrm{M}\sim 4 \text{ nm}}$, is much smaller than the separation distance of the electrodes, ${L\sim 1 \text{ \textmu m}}$, in physiological settings.
A voltage of $2V$ is suddenly applied to the system and the transient behavior of the membrane is studied until it relaxes to a new equilibrium. The dielectric constants of the electrolyte and the membrane are assumed to be ${\epsilon\approx 80\epsilon_0}$ and ${\epsilon^{\rm M}\approx 4\epsilon_0}$, respectively, with $\epsilon_0$ being the vacuum permittivity. This leads to a dielectric mismatch of ${\Gamma=\epsilon/\epsilon^{\rm M}\approx 20}$.
Due to its low dielectric permittivity, we assume that the membrane is impermeable to ionic species. Given the planar symmetry of the problem, all spatial variations occur along the direction perpendicular to the membrane and the electrode surfaces.

\subsection{Governing Equations}
In the absence of advection, the spatiotemporal dynamics of the electrolyte solutions can be described by the  Poisson-Nernst-Planck equations~\cite{Nernst1888,Nernst1889, Planck1890}, which arise as the dilute limit of the Onsager transport theory~\cite{fong2020transport} and are given by
\begin{subequations}
\label{eq:PNP-dim}
    \begin{align}
        \label{eq:PNP-C}
        \frac{\partial C_i^\alpha}{\partial t} &= - \frac{\partial j_i^\alpha}{\partial x} \ , \\
        \label{eq:PNP-J}
        j_i^\alpha &= - D_i \left( \frac{\partial C_i^\alpha}{\partial x} + \frac{\mathrm{z}_i \mathrm{e} C_i^\alpha}{k_{\rm B}T} \frac{\partial \phi^\alpha}{\partial x} \right) \ , \\
        \label{eq:PNP-Phi}
        -\epsilon \frac{\partial^2 \phi^\alpha}{\partial x^2} &= \sum_i \mathrm{z}_i \mathrm{e} C_i^\alpha = \rho^\alpha \ , \\
        \label{eq:PNP-PhiM}
        -\epsilon^{\rm M}\frac{\partial^2 \phi^\mathrm{M}}{\partial x^2} &= 0 \ .
    \end{align}
\end{subequations}
Here, ${C_i^\alpha \left(x,t\right)}$ is the concentration of ionic species $i$ at location $x$ and time $t$ within domain ${\alpha\in\{\mathrm{I},\mathrm{II}\}}$ . Equation~\eqref{eq:PNP-C} is the mass balance for each species and Eq.~\eqref{eq:PNP-J} is the constitutive relation for the flux ${j_i^\alpha \left(x,t\right)}$ consisting of diffusion and electromigration terms, with $D_i$ being the diffusion coefficient of each species, $\mathrm{z}_i$ the ion valence, $\mathrm{e}$ the fundamental charge, and $k_{\rm B}T$ the thermal energy scale.
Equations~\eqref{eq:PNP-Phi} and~\eqref{eq:PNP-PhiM} are Gauss's law in the electrolyte and membrane, respectively, where we assume the concentration of ionic species to be negligible inside the membrane due to the low dielectric permittivity of the lipid membrane.
We consider a symmetric binary monovalent electrolyte, where ${i \in \{+,-\}}$ and ${\mathrm{z}_+ = -\mathrm{z}_- = 1}$. We also assume equal diffusivity for both ions, i.e., ${D_i = D}$.

\subsection{Initial and Boundary Conditions}

The binary electrolyte solutions are assumed initially to be homogeneous and symmetric, i.e., ${C_\pm^\alpha \left(x,0\right) = C_0}$. For both species, we impose no-flux boundary conditions, i.e., ${j_\pm^\alpha = 0}$ at both the electrode and membrane surfaces. Electrostatics requires that the electric potential is continuous at the membrane-electrolyte interfaces, i.e., ${\phi^\alpha = \phi^\mathrm{M}}$.
Because the membrane is assumed to carry no surface charge, the electric displacement field is also continuous at the membrane-electrolyte interfaces, i.e., ${\epsilon \, \partial\phi^\alpha / \partial x = \epsilon^\mathrm{M} \, \partial\phi^\mathrm{M} / \partial x}$.
We choose a potential reference ${\phi^\alpha = \mp\, V}$ at the electrode surfaces, such that there is an applied potential difference of $2V$.
Here, $-$ and $+$ correspond to ${\alpha={\rm I}}$ and ${\alpha={\rm II}}$, respectively.

\begin{figure*}
    \centering
    {
        \hfill
        \includegraphics[width=0.495\linewidth]{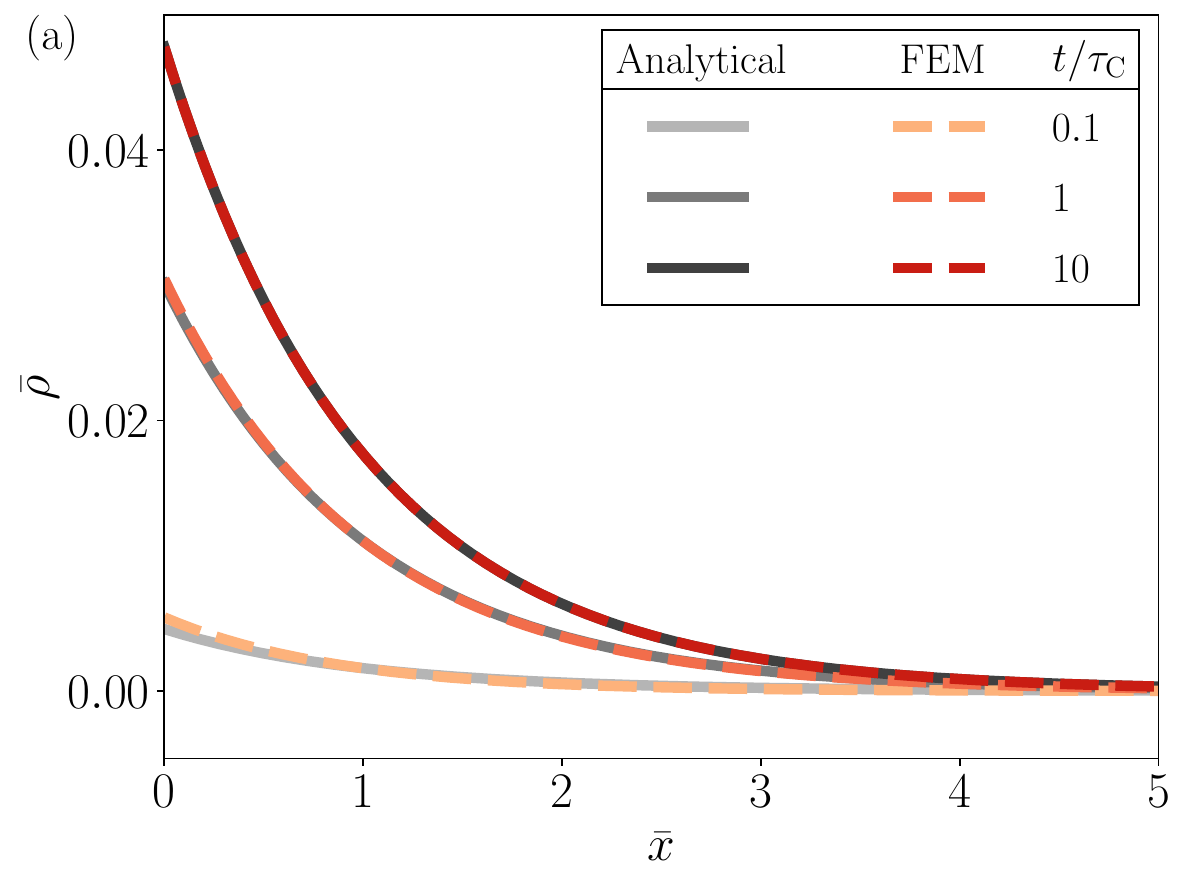}
        \hfill
        \includegraphics[width=0.495\linewidth]{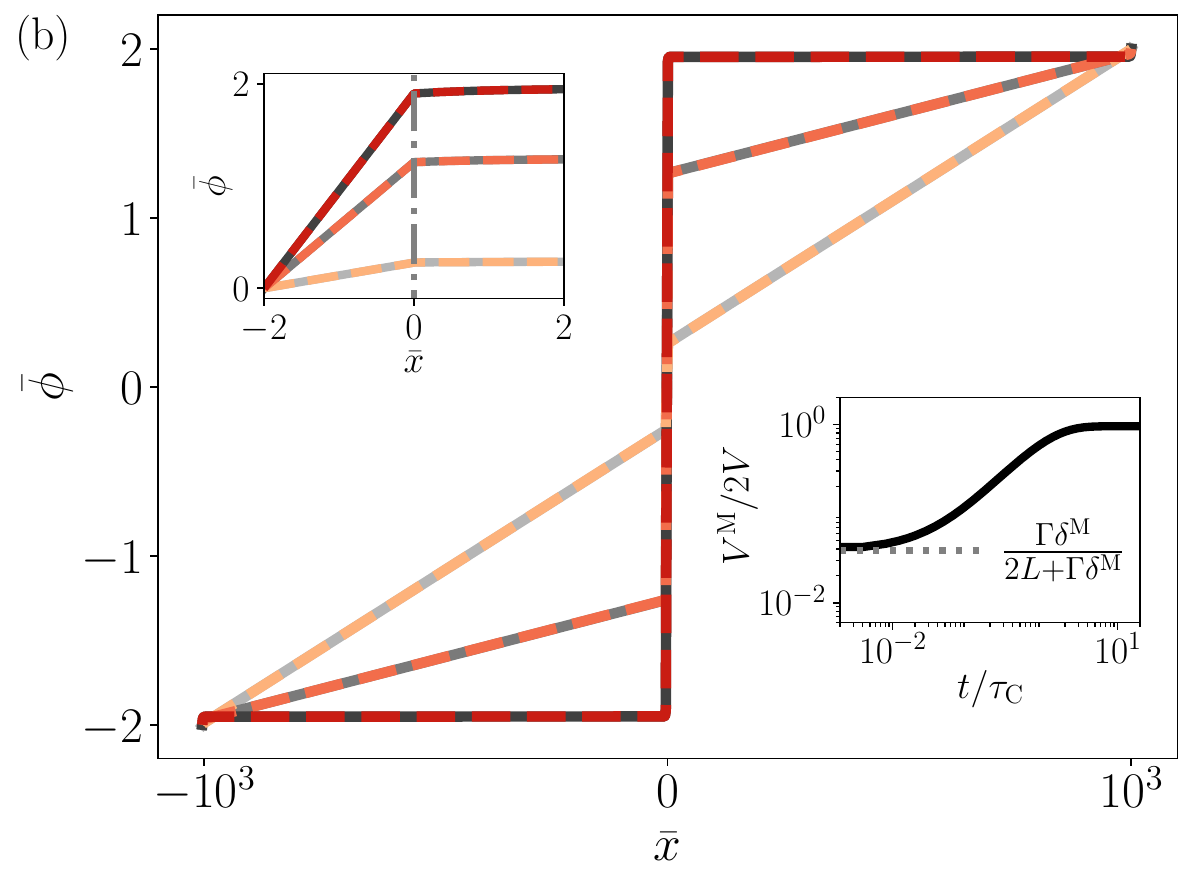}
        \hfill
    }
    \vspace{-0.5cm}
    \caption{The characteristic timescale for the membrane-electrolyte system to respond to the applied potential is $\tau_{\rm C}$, given by Eq.~\eqref{eq:dim-tauC}. We show the system response for ${t/\tau_{\rm C}=0.1,1,10}$.
    Panel (a) shows the formation of the diffuse charge layer through the charge density at the membrane-electrolyte interface in domain II and panel (b) shows the potential profile across the entire system. In (b), the upper left inset highlights the potential profile near the membrane-electrolyte interface at $\bar{x}=0$ (dashed gray line). The lower right inset shows the transmembrane potential, $V^{\rm M}$, evolving from its initial value to equilibrium as the membrane is charged. By ${t=10\tau_{\rm C}}$, we find that the system has approximately reached the equilibrium profile. For these plots, we take the parameters to be ${\bar{L} = 10^3}$, ${\Gamma\bar{\delta}^{\rm M} = 80}$, and ${\bar{V} = 2}$. The analytical solution plotted is the sum of the first- and third-order contributions from the perturbation analysis, which will be introduced later.
    }
    \label{fig:phi-rho}
\end{figure*}

\subsection{Nondimensionalization}

The governing equations and boundary conditions are nondimensionalized by the Debye screening length, ${\lambda_{\rm D} = \sqrt{\epsilon k_{\rm B} T/2 \mathrm{e}^2 C_0}}$, which characterizes the thickness of the diffuse charge layer; the Debye time, ${\tau_{\rm D} = \lambda_{\rm D}^2/D}$, which sets the diffusion timescale over the diffuse charge layer; the thermal voltage, ${\phi_{\rm T} = k_{\rm B} T/\mathrm{e}}$; and the total ion concentration, $2C_0$. For a binary monovalent electrolyte, the problem can be conveniently analyzed in terms of the overall salt concentration and the charge density~\cite{Bazant04, row2025spatiotemporal}.
Accordingly, we introduce the dimensionless variables ${\bar{c}^\alpha = \left(C_+^\alpha + C_-^\alpha\right) / 2C_0}$ and ${\bar{\rho}^\alpha = \left(C_+^\alpha - C_-^\alpha\right) / 2C_0}$, respectively, which can be used to recast the system of Eqs.~\eqref{eq:PNP-dim} as
\begin{subequations}
\label{eq:PNP-nondim}
    \begin{align}
        \label{eq:ND-c}
        \frac{\partial \bar{c}^\alpha}{\partial \bar{t}} &= \frac{\partial}{\partial \bar{x}} \left( \frac{\partial \bar{c}^\alpha}{\partial \bar{x}} + \bar{\rho}^\alpha \frac{\partial \bar{\phi}^\alpha}{\partial \bar{x}} \right) \ , \\
        \label{eq:ND-rho}
        \frac{\partial \bar{\rho}^\alpha}{\partial \bar{t}} &= \frac{\partial}{\partial \bar{x}} \left( \frac{\partial \bar{\rho}^\alpha}{\partial \bar{x}} + \bar{c}^\alpha \frac{\partial \bar{\phi}^\alpha}{\partial \bar{x}} \right) \ , \\
        \label{eq:ND-phi}
        \frac{\partial^2 \bar{\phi}^\alpha}{\partial \bar{x}^2} &= - \bar{\rho}^\alpha \ , \\
        \label{eq:ND-phiM}
        \frac{\partial^2 \bar{\phi}^\mathrm{M}}{\partial \bar{x}^2} &= 0 \ ,
    \end{align}
\end{subequations}
where ${\bar{t} = t/\tau_{\rm D}}$, ${\bar{x} = x/\lambda_{\rm D}}$, and ${\bar{\phi} = \phi / \phi_{\rm T}}$.

The initial and boundary conditions can also be non-dimensionalized.
The electrolyte solutions are initially homogeneous, so that ${\bar{c}^\alpha \left(\bar{x},0\right) = 1}$ and ${\bar{\rho}^\alpha \left(\bar{x},0\right) = 0}$.
At all interfaces, there should no flux of each species, leading to
\begin{subequations}
    \begin{align}
        \label{eq:BC-c}
        \frac{\partial \bar{c}^\alpha}{\partial \bar{x}} + \bar{\rho}^\alpha \frac{\partial \bar{\phi}^\alpha}{\partial \bar{x}} &= 0 \ , \\
        \label{eq:BC-rho}
        \frac{\partial \bar{\rho}^\alpha}{\partial \bar{x}} + \bar{c}^\alpha \frac{\partial \bar{\phi}^\alpha}{\partial \bar{x}} &= 0 \ ,
    \end{align}
     and the potential should match the external voltage at the electrode surfaces,
    \begin{equation}
        \label{eq:BC-phi}
        \bar{\phi}^\alpha = \mp\, \bar{V} \ ,
    \end{equation}
    where ${\bar{V} = V / \phi_{\rm T}}$.
    At the membrane-electrolyte interfaces, the boundary conditions are
    \begin{align}
        \label{eq:BC-phiM}
        \bar{\phi}^\alpha &= \bar{\phi}^\mathrm{M} \ , \\
        \label{eq:BC-dphi}
        \frac{\partial \bar{\phi}^\alpha}{\partial \bar{x}} &= \frac{1}{\Gamma} \frac{\partial \bar{\phi}^\mathrm{M}}{\partial \bar{x}} \ ,
    \end{align}
    with $\Gamma$ being the dielectric mismatch.
    
    For the problem setup in Fig.~\ref{fig:fig_schematic}, the governing equations and boundary conditions are symmetric with respect to the membrane centerline, so that $\bar{\rho}$ and $\bar{\phi}$ are odd and $\bar{c}$ is even.
    Therefore, ${\bar{\phi}^\mathrm{M}=0}$ at the center, so Eq.~\eqref{eq:ND-phiM} together with Eqs.~\eqref{eq:BC-phiM} and~\eqref{eq:BC-dphi} imply a Robin-type boundary condition at the membrane surfaces
    \begin{equation}
        \label{eq:BC-Robin}
        \frac{\partial \bar{\phi}^\alpha}{\partial \bar{x}} = \mp \, \frac{2}{\Gamma\bar{\delta}^{\rm M}} \, \bar{\phi}^\alpha \ ,
    \end{equation}
\end{subequations}
where ${\bar{\delta}^{\rm M} = \delta^\mathrm{M} / \lambda_{\rm D}}$ is the dimensionless membrane thickness.
We recognize that ${\Gamma\bar{\delta}^{\rm M} = \left(\epsilon/\lambda_{\rm D}\right)/\left(\epsilon^{\rm M}/\delta^{\rm M}\right)}$ is the ratio of the diffuse charge layer capacitance to the membrane capacitance.

Equation~\eqref{eq:BC-Robin} decouples the three domains, allowing us to focus exclusively on the electrolyte in domain II, and subsequently reconstruct the full solution across all three domains using the symmetry of $\bar{\rho}$, $\bar{\phi}$, and $\bar{c}$. For convenience, we set ${\bar{x}=0}$ at the membrane-interface interface and solve Eqs.~\eqref{eq:ND-c}--\eqref{eq:ND-phi} over the interval ${\left[0, \, \bar{L}\right]}$, where ${\bar{L}=L/\lambda_{\rm D}}$. We also drop the superscript II from this point onward as we only consider domain II for the electrolyte.

\begin{figure}
    \centering
        \includegraphics[width=0.97\linewidth]{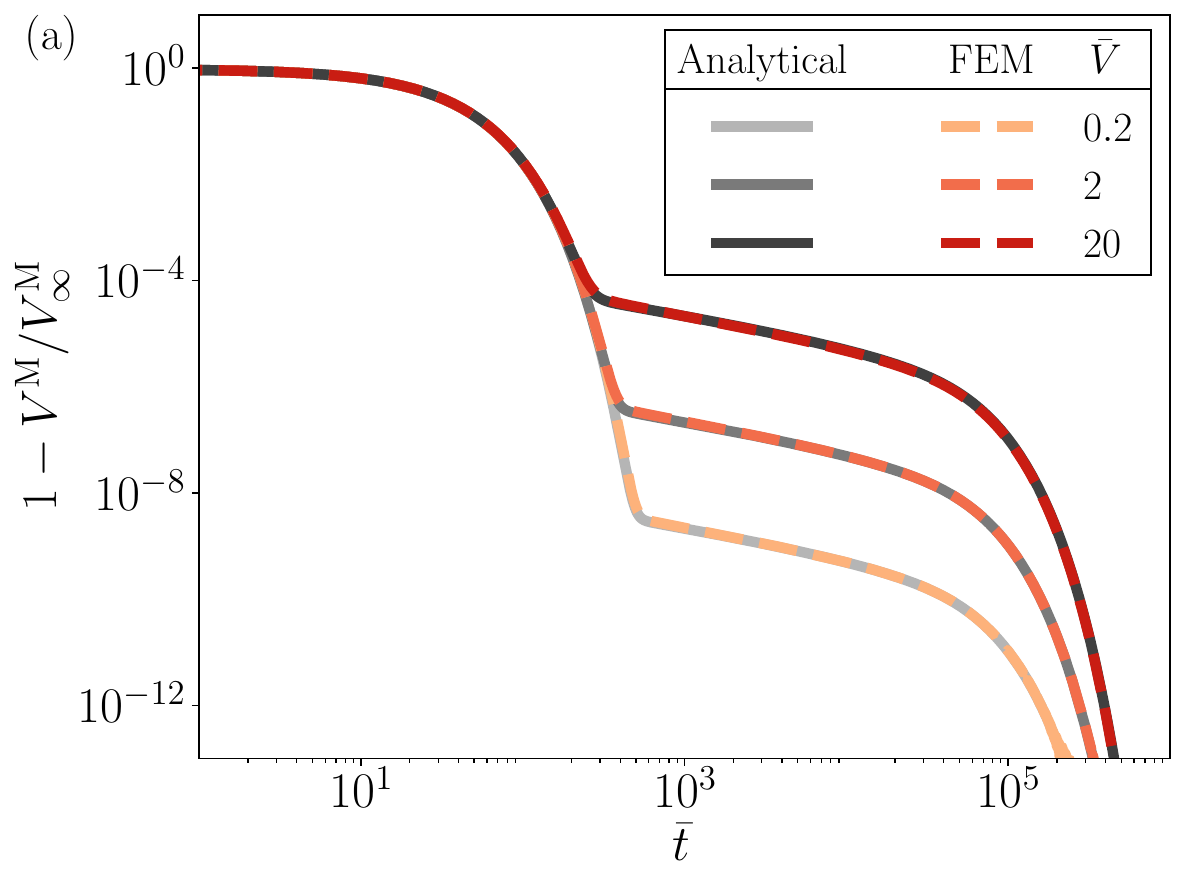}
        \includegraphics[width=0.97\linewidth]{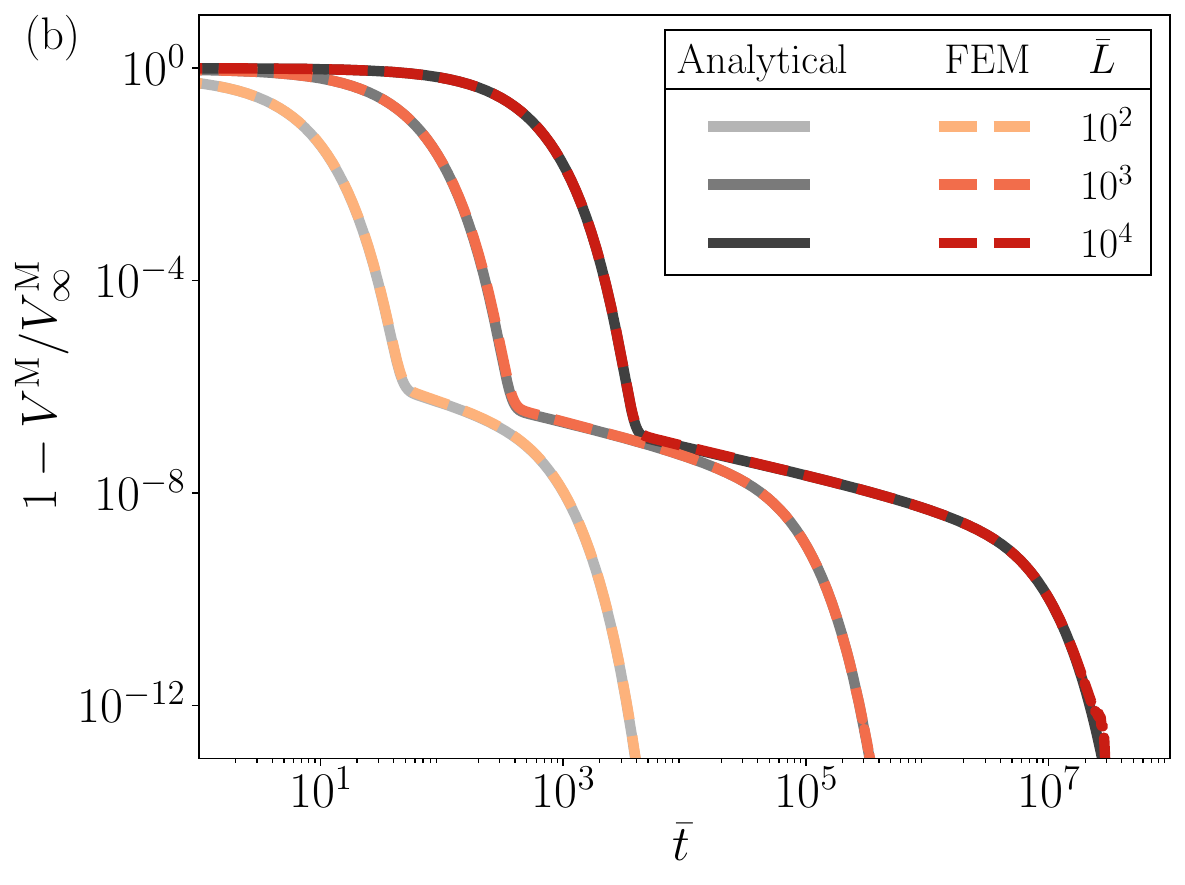}
        \includegraphics[width=0.97\linewidth]{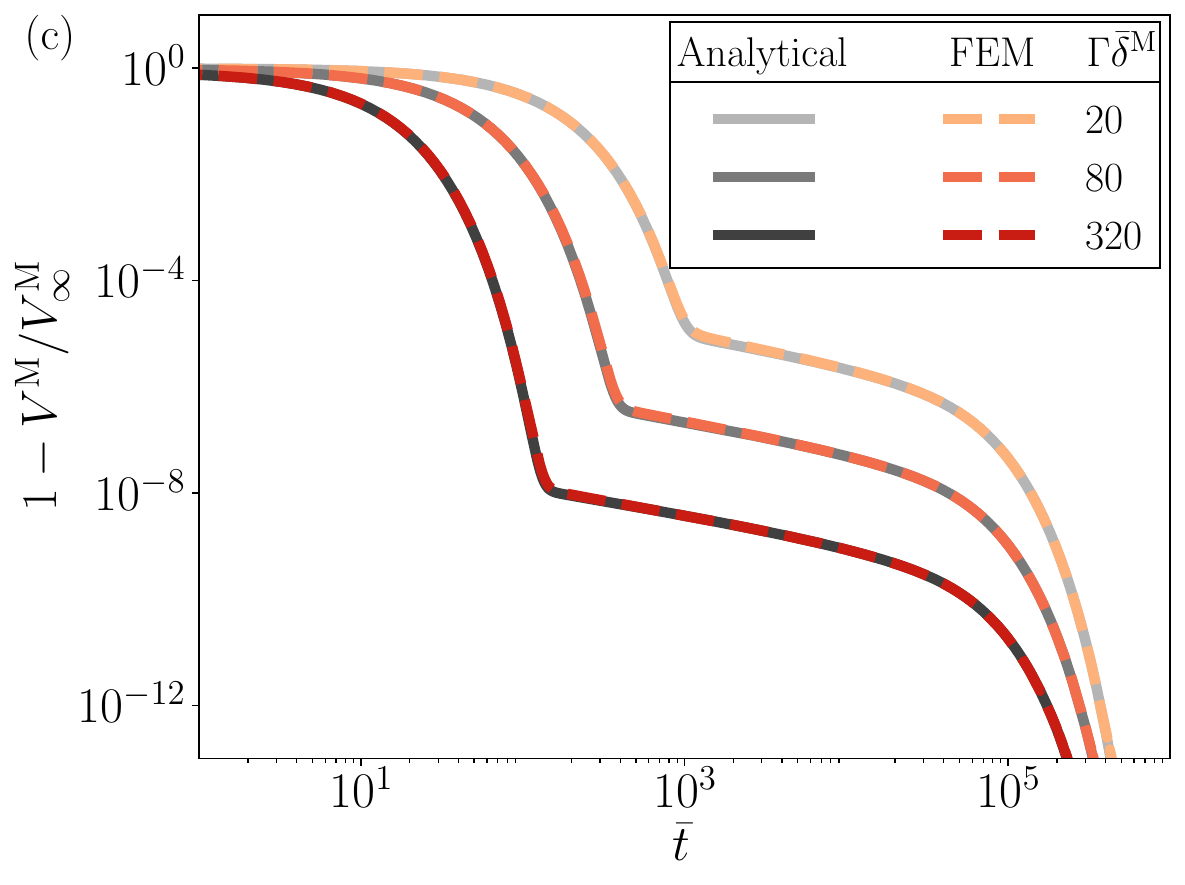}
    \caption{Panels (a)--(c) depict approach of the transmembrane potential, ${V^{\rm M}\left(t\right)}$ to its equilibrium value, $V^{\rm M}_{\infty}$ for different values of model parameters. Since $V^{\rm M}$ grows monotonically, we plot the approach as the difference between 1 and the fraction of the equilibrium value achieved. The base values we choose are ${\bar{V}=2}$ (${V\sim 50\text{ mV}}$) for the applied potential, ${\bar{L}=10^3}$ (${L\sim 1 \text{ \textmu m}}$) for the system size, and ${\Gamma\bar{\delta}^{\rm M}=80}$ for the capacitance ratio. The panels show the effect of varying (a) $\bar{V}$, (b) $\bar{L}$, and (c) $\Gamma\bar{\delta}^{\rm M}$. In all cases, we observe a two-step charging process, each relaxing over distinct timescales. The analytical solution is the transmembrane potential up to third order from Eq.~\eqref{eq:deltav-nonlinear}.}
    \label{fig:transmembrane}
\end{figure}

\section{Numerical Observations}
\label{sec:numerics}
We first present the numerical results obtained from solving the system of coupled nonlinear partial differential equations, Eqs.~\eqref{eq:PNP-nondim}, in the entire domain, i.e.~$\{\mathrm{I}, \mathrm{M}, \mathrm{II}\}$, obtained using the nonlinear finite element method~\cite{papadopoulos2015fem, hughes2012finite}. Details on how the spatial extent of the diffuse charge layers is resolved and how adaptive time stepping is implemented can be found in the supplementary material of Ref.~\cite{row2025spatiotemporal}.

Figures~\ref{fig:phi-rho}(a) and (b) show the spatial profiles of the charge density and the electric potential, respectively, at three representative time points for a physiological set of parameters. Note that time is rescaled by the capacitive timescale $\tau_{\rm C}$ from Eq.~\eqref{eq:dim-tauC}, which will shortly be shown to serve as the primary relaxation timescale. As expected, the charge density is nonzero only within a few Debye lengths near the membrane interface (also at the electrode surface, not shown) and decays exponentially to zero with distance from the surface, consistent with the formation of diffuse charge layers on either side of the membrane.
In Fig.~\ref{fig:phi-rho}(b), the electric potential varies linearly through the majority of the bulk electrolyte where ${\bar{\rho}\approx 0}$. The potential varies linearly in the membrane and saturates exponentially across the diffuse charge layer.  At long times, the profile converges to an equilibrium solution in which the potential is approximately constant in the bulk electrolyte while most of the potential difference occurs over the membrane.

A quantity of interest is the transmembrane potential $\bar{V}^{\rm M}$, defined as
\begin{equation}
    \bar{V}^{\rm M}\left(\bar{t}\right)=\bar{\phi}^{\rm M}\left(0,\bar{t}\right)-\bar{\phi}^{\rm M}\left(-\bar{\delta}^{\rm M},\bar{t}\right)=2\bar{\phi}\left(0,\bar{t}\right)\ , \label{eq:def-transmembrane}
\end{equation}
which grows monotonically from its initial nonzero value to an equilibrium value, which we denote $V^{\rm M}_\infty$, as shown in the Fig.~\ref{fig:phi-rho}(b) lower right inset.
Figure~\ref{fig:transmembrane} plots the approach of ${V^{\rm M}\left(t\right)}$ to its equilibrium value for different values of the applied voltage $\bar{V}$, system size $\bar{L}$, and capacitance ratio $\Gamma\bar{\delta}^{\rm M}$. 
In all cases, the membrane potential develops through two distinct exponential charging processes, which we refer to as the initial and later regimes, respectively. Furthermore, increasing the applied voltage as in Fig.~\ref{fig:transmembrane}(a) leaves the initial regime unchanged but causes the secondary regime to begin earlier indicating the later process is nonlinear with respect to the applied voltage. Increasing the system size $\bar{L}$, as in Fig.~\ref{fig:transmembrane}(b), does not appreciably impact the response magnitude in either regime, but the timescales of both increase, with a more pronounced effect in the later regime. 
Finally, varying the membrane properties through $\Gamma\bar{\delta}^{\rm M}$ as in Fig.~\ref{fig:transmembrane}(c) alters the timescale of the initial regime and notably changes the response magnitude of the later regime, suggesting that this parameter affects both processes and contributes significantly to the nonlinearity. 

\section{Analytical Results}
\label{sec:results}

\subsection{Perturbation Expansion}
From the behaviors observed in Fig.~\ref{fig:transmembrane}, two distinct processes control the relaxation of the transmembrane potential and these operate on disparate timescales.
This motivates representing the solution as a superposition of two contributions, each with its own timescale.
Moreover, the observation in Fig.~\ref{fig:transmembrane}(a) that the initial response is independent of $\bar{V}$ while the magnitude of the secondary response increases with $\bar{V}$, suggests a pertubation expansion in $\bar{V}$ may be appropriate.
The leading-order linear dynamics in $\bar{V}$ govern the initial relaxation, whereas the long-time relaxation is driven by higher-order dynamics in $\bar{V}$.

However, the perturbation analysis must also account for the role of the capacitance mismatch ($\Gamma\bar{\delta}^{\rm M}$) in driving the nonlinear response~(Fig.~\ref{fig:transmembrane}(c)), indicating that effects from $\Gamma \delta^{\mathrm{M}}$ must also be included in the perturbation parameter. Mathematically, the nonlinearity is driven by the electromigration terms (see Eq.~\eqref{eq:PNP-J}), which contribute only in regions where
the charge density is nonzero or where concentration gradients are subject to electric fields.
This occurs exclusively within the diffuse charge layers. Accordingly, one may propose that the appropriate perturbation parameter must be the total charge per unit area in each diffuse layer. 
This quantity can be readily estimated from the steady-state equivalent circuit presented in Ref.~\cite{row2025spatiotemporal} for a charging membrane. Given the diffuse layer capacitance ${C_{\rm D}=\epsilon/\lambda_{\rm D}}$, the membrane capacitance ${C_{\rm M}=\epsilon^{\rm M}/\delta^{\rm M}}$, and noting that there are four diffuse layers (one on each surface, including electrodes) in series with the membrane, the total capacitance is given by ${C_{\text{tot} }/C_\text{D}=1/\left(4+\Gamma\bar{\delta}^{\rm M}\right)}$. The dimensionless charge in each diffuse layer is then ${\eta = 2\bar{V}C_\text{tot}/C_{\rm D} = 2\bar{V}/\left(4+\Gamma\bar{\delta}^{\rm M}\right)}$, which serves as the perturbation parameter.

The linear analysis presented in the following two sections will indeed confirm that $C_\text{tot}$ and $\eta$ represent the total capacitance and charge on each capacitor, respectively. With these definitions, it is readily seen that $\eta$ may remain small even for large applied voltages relative to the thermal voltage, since ${\Gamma\bar{\delta}^{\rm M}\approx 100 \gg 1}$ in physiological settings.
For instance, ${\eta\approx 1}$ corresponds to an applied voltage of approximately ${2.5\, \rm{V}}$, which suggests that the nonlinear analysis is applicable to the full range of physiologically-relevant transmembrane potentials.
Given the perturbation parameter, we may then expand the relevant fields as 
\begin{subequations}
\label{eq:perturb}
    \begin{align}
        \label{eq:perturb-c}
        \bar{c} \left(\bar{x},\bar{t};\eta\right) = \sum_{i=0}^\infty \eta^i \bar{c}_i\left(\bar{x},\bar{t}\right) &= \bar{c}_0+\eta^2\bar{c}_2+\cdots \ , \\
        \label{eq:perturb-rho}
        \bar{\rho} \left(\bar{x},\bar{t};\eta\right) = \sum_{i=0}^\infty \eta^i \bar{\rho}_i\left(\bar{x},\bar{t}\right) &= \eta\bar{\rho}_1+\eta^3\bar{\rho}_3+\cdots \ , \\
        \label{eq:perturb-phi}
        \bar{\phi} \left(\bar{x},\bar{t};\eta\right) = \sum_{i=0}^\infty \eta^i \bar{\phi}_i\left(\bar{x},\bar{t}\right) &= \eta\bar{\phi}_1+\eta^3\bar{\phi}_3+\cdots\ .
    \end{align}
\end{subequations}
Here, the expansions for $\bar{\rho}$ and $\bar{\phi}$ only contain odd powers of $\eta$ due to their odd symmetry, while $\bar{c}$ only contains even powers due to its even symmetry. 
These expansions can now be substituted into the governing equations and boundary conditions to yield the response at each perturbation order, which we now analyze.

\subsection{Linear Dynamics}
The linear regime response depends only on the zeroth- and first-order solutions in the perturbation expansion. 

$\\$
\textbf{Zeroth Order:}
The zeroth-order solution ($\eta^0$) is trivial and corresponds to ${V = 0}$, where the system remains at its initial condition. This amounts to saying,
\begin{equation}
    \label{eq:c0}
    \bar{c}_0\left(\bar{x},\bar{t}\right) = 1 \ ,
\end{equation}
for all locations and times.

$\\$
\textbf{First Order:} The first-order solution ($\eta^1$) can be obtained by solving the following governing equations
\begin{subequations}
    \label{eq:first-order}
    \begin{align}
        \label{eq:rho1}
        \frac{\partial \bar{\rho}_1}{\partial \bar{t}} &= \frac{\partial^2 \bar{\rho}_1}{\partial \bar{x}^2} - \bar{\rho}_1 \ , \\
        \label{eq:phi1}
        \frac{\partial^2 \bar{\phi}_1}{\partial \bar{x}^2} &= - \bar{\rho}_1 \ ,
    \end{align}
\end{subequations}
subjected to the boundary conditions
\begin{subequations}
\label{eq:first-order-BCs}
    \begin{align}
        \label{eq:BC-rho1}
        \left[ \frac{\partial \bar{\rho}_1}{\partial \bar{x}} + \frac{\partial \bar{\phi}_1}{\partial \bar{x}} \right]_{\bar{x}=0,\bar{L}} &= 0 \ , \\
        \label{eq:BC-Robin1}
        \left.\frac{\partial \bar{\phi}_1}{\partial \bar{x}} \right|_{\bar{x}=0} &= \frac{2}{\Gamma\bar{\delta}^{\rm M}} \, \bar{\phi}_1 \left(0,\bar{t}\right) \ , \\
        \label{eq:BC-phi1}
        \bar{\phi}_1 \left(\bar{L}, \bar{t}\right) &= \frac{1}{2} \left(4 + \Gamma\bar{\delta}^{\rm M}\right) \ ,
    \end{align}
\end{subequations}
and initial condition ${\bar{\rho}_1 \left(\bar{x},0\right) = 0}$.

Equations~\eqref{eq:rho1}--\eqref{eq:BC-phi1} can be solved by the method of separation of variables to yield long-time solutions for $\bar{\rho}_1$ and $\bar{\phi}_1$ (Appendix~\ref{app:firstorder}). Assuming, ${\bar{L} \gg 1}$, such that ${e^{-\bar{L}} \approx 0}$, the long-time  solution for ${\bar{t} \gg 1}$ is given by 
\begin{subequations}\label{eq:linear-sol}
    \begin{align}
        \label{eq:rho1-t}
        \bar{\rho}_1 \left(\bar{x},\bar{t}\right) &= f\left(\bar{x}\right) \left( 1 - e^{-\bar{t}/\bar{\tau}_\mathrm{C}} \right)\ , \\
        \label{eq:phi1-t}
        \begin{split}
            \bar{\phi}_1 \left(\bar{x}, \bar{t}\right) &=
            \frac{1}{2}\left(4 + \Gamma\bar{\delta}^{\rm M}\right) + \left(\frac{\bar{x}-\bar{L}}{\bar{\tau}_{\rm C}}\right)e^{-\bar{t}/\bar{\tau}_\mathrm{C}} \\
            &\qquad\, - \left[ f\left(\bar{x}\right) + 1 \right] \left( 1 - e^{-\bar{t}/\bar{\tau}_\mathrm{C}} \right)\ ,
        \end{split}
    \end{align}
    where
    \begin{align}
        \label{eq:f}
        f\left(\bar{x}\right) &= e^{-\bar{x}} - e^{-(\bar{L}-\bar{x})}\ , \\
        \label{eq:tM}
        \bar{\tau}_\mathrm{C} &= \frac{2\bar{L} + \Gamma\bar{\delta}^{\rm M}}{4 + \Gamma\bar{\delta}^{\rm M}} \ .
    \end{align}
\end{subequations}
The function ${f\left(\bar{x}\right)}$ shows the formation of two boundary layers---one at the membrane interface and one at the electrode interface---each with a characteristic thickness $\lambda_{\rm D}$ and carrying equal and opposite charges. As seen from Eq.~\eqref{eq:rho1-t}, these layers develop on the timescale $\bar{\tau}_\mathrm{C}$. The electric potential in Eq.~\eqref{eq:phi1-t} exhibits a similar structure to that of the charge density with an additional linear contribution that relaxes to a constant equilibrium value over the same timescale $\bar{\tau}_\mathrm{C}$, as in Fig.~\ref{fig:phi-rho}. See Appendix~\ref{app:firstorder} for additional details on the analytical solution.\footnote{Note that the solution presented in Eqs.~\eqref{eq:linear-sol} do not satisfy Eqs.~\eqref{eq:rho1} and~\eqref{eq:BC-rho1} exactly at short times due to the assumption ${\bar{t}\gg 1}$ (i.e., ${t\gg \tau_{\rm D}}$).
At these short times, the diffuse charge layers are still forming, as shown in \cite{row2025spatiotemporal}, and the exponential profile does not yet hold.
Therefore, this analysis is only valid for ${\bar{t}\gg 1}$, i.e., on timescales much larger than the Debye timescale.}

The initial capacitive response of biological membranes can be understood through the relaxation of the transmembrane potential. To that end, from Eq.~\eqref{eq:def-transmembrane} and Eq.~\eqref{eq:phi1-t}, the linear contribution to the dimensional transmembrane potential is given by
\begin{equation}
    \label{eq:deltaV-tau}
    V^\mathrm{M}_\text{lin} \left(t\right) = \frac{2V}{1 + 4/\Gamma\bar{\delta}^{\rm M}} \left( 1 - \frac{\bar{L}-2}{\bar{L}+\Gamma\bar{\delta}^{\rm M}/2} \, e^{-t/\tau_\mathrm{C}} \right)\ ,
\end{equation}
where ${\tau_\mathrm{C} = \bar{\tau}_\mathrm{C} \tau_{\rm D}}$ is the dimensional membrane relaxation timescale.
Equation~\eqref{eq:deltaV-tau} indicates that at time ${t=0}$, the membrane immediately senses a fraction ${\Gamma\bar{\delta}^{\rm M}/\left(2\bar{L}+\Gamma\bar{\delta}^{\rm M}\right)}$ of the applied potential, which becomes negligible for large ${\bar{L}\gg \Gamma \bar{\delta}^{\rm M}}$. 
For times ${t \gg\tau_\mathrm{C}}$, the linear contribution to the transmembrane potential corresponds to a fraction ${1/\left(1+4/\Gamma\bar{\delta}^\mathrm{M}\right)}$ of the applied potential. In biological membranes, where ${\Gamma \bar{\delta} \approx 100}$ for physiological concentrations of approximately ${150 \text{ mM}}$, this effectively amounts to the entire applied potential. 

The linear regime relaxation timescale can be compared with the RC timescale analyzed by Bazant \textit{et al.}~\cite{Bazant04} for a bare electrolyte without the membrane. Rewriting Eqs.~\eqref{eq:dim-tauC} or \eqref{eq:tM} as 
\begin{equation}
    \label{eq:tauM}
    \tau_\mathrm{C} = \frac{\tau_\mathrm{B}}{2 + \Gamma\bar{\delta}^{\rm M}/2} + \frac{\tau_{\rm D}}{1 + 4 / \Gamma\bar{\delta}^{\rm M}}\ ,
\end{equation}
where ${\tau_\mathrm{B} = \lambda_{\rm D}L/D}$ is the bare electrolyte timescale, it can be seen that the capacitive response of the membrane-electrolyte system is significantly faster than that of a bare electrolyte, i.e. ${\tau_\mathrm{C} \ll \tau_\mathrm{B}}$. 
Furthermore, for large $\Gamma\bar{\delta}^{\rm M}$, and a system size $\bar{L}$ on the order of $\Gamma\bar{\delta}^{\rm M}$, we observe that the capacitive timescale, $\tau_{\rm C}$, becomes of the same order as the Debye timescale, $\tau_{\rm D}$.
That is, for applied potential differences on length scales of ${100 \text{ nm}}$, the relaxation dynamics become ultrafast, i.e., ${\sim 1\text{ ns}}$, wherein the adjustment of ions in the diffuse layers and the relaxation of the transmembrane potential become comparable.
Finally, Fig.~\ref{fig:transmembrane} shows agreement between the initial regime and the analytical results from linear-order contributions for all parameters.

\subsection{Equivalent Circuit}

\begin{figure}[t]
    \centering
    \includegraphics[width=\linewidth]{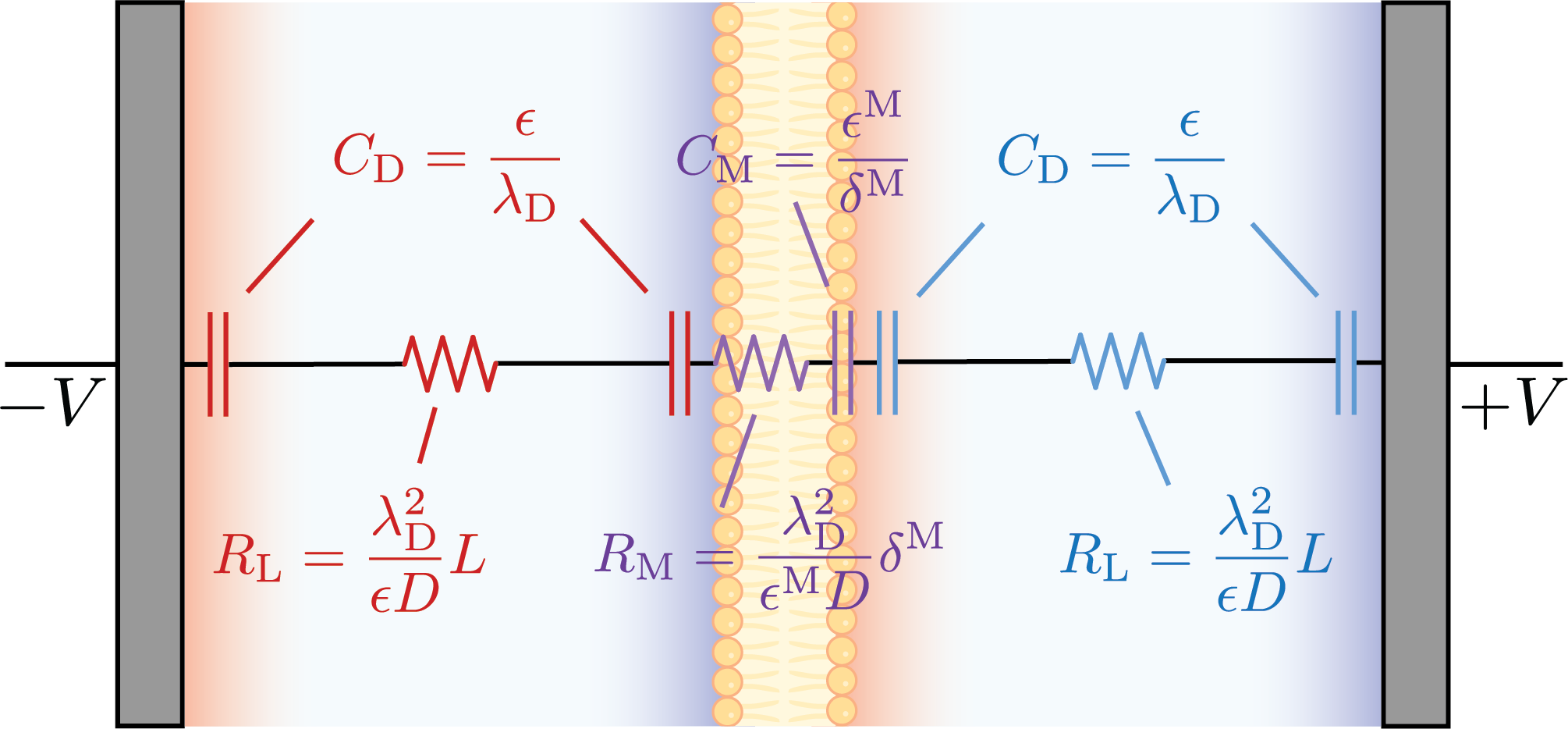}
    \caption{The equivalent circuit representation for the membrane-electrolyte charging problem in Fig.~\ref{fig:fig_schematic}. Each electrolyte consists of a bulk resistance and two diffuse charge layer capacitance contributions. The membrane consists of an effective capacitance and an apparent resistance that arises from continuity of the electric displacement field.}
    \label{fig:equiv-circuit}
\end{figure}
The initial relaxation process can be understood by an equivalent circuit. Consider the enclosed charge in the diffuse charge layer around the membrane. For thin diffuse charge layers relative to electrolyte thickness (${\lambda_{\rm D}\ll L}$), the enclosed charge can be calculated by considering the region from the membrane surface to the center of the electrolyte domain. Introducing the dimensionless charge per unit area as ${\bar{Q} = Q/\left( \epsilon \phi_{\rm T}/\lambda_{\rm D}\right)}$, Gauss's law yields
\begin{equation}
    \label{eq:Gauss-non}
    \bar{Q} = \left( -\frac{\partial \bar{\phi}}{\partial \bar{x}} \right)_{\bar{x} = \bar{L}/2} - \left( -\frac{\partial \bar{\phi}}{\partial \bar{x}} \right)_{\bar{x} = 0} \ .
\end{equation}
To linear order, where ${\bar{\phi} = \eta \bar{\phi}_1 = 2\bar{V}\bar{\phi}_1/\left(4+\Gamma\bar{\delta}^{\rm M}\right)}$, the total enclosed charge per unit area is then 
\begin{equation}
    \label{eq:q}
    \bar{Q} \left(\bar{V},\bar{t}\right) = \frac{2 \bar{V}}{4 + \Gamma\bar{\delta}^{\rm M}} \left( 1 - e^{-\bar{t}/\bar{\tau}_\mathrm{C}} \right)\ .
\end{equation}
From this expression, the total dimensionless capacitance may be evaluated as 
\begin{equation}
    \label{eq:capacitance}
    \left.\frac{\partial \bar{Q}}{\partial \left(2\bar{V}\right)} \right|_{\bar{t} \to \infty} = \frac{1}{4 + \Gamma\bar{\delta}^{\rm M}} = \bar{C}_\text{tot} \ ,
\end{equation}
confirming that $\bar{C}_\text{tot}$ indeed represents the system capacitance, as indicated earlier. In dimensional units, the capacitance can be expressed as 
\begin{equation}
    \label{eq:capacitance-inverse}
    \frac{1}{C_\text{tot}} = \frac{4}{C_{\rm D}} + \frac{1}{C_{\rm M}} \ ,
\end{equation}
where ${C_{\rm D} = \epsilon / \lambda_{\rm D}}$, ${C_{\rm M} = \epsilon^\mathrm{M} / \delta^\mathrm{M}}$, and ${\bar{C}_\text{tot} = C_\text{tot}/C_{\rm D}}$.
Equation~\eqref{eq:capacitance-inverse} implies a circuit of five capacitors in series. These include four capacitors corresponding to the four diffuse charge layers (one at each electrode and one at each membrane surface), each with capacitance $C_{\rm D}$, and an additional capacitor corresponding to the membrane with capacitance $C_{\rm M}$ (see Fig.~\ref{fig:equiv-circuit}).
It is worth noting that in physiological settings, the ratio ${C_{\rm M} / C_{\rm D} = 1 / \Gamma \bar{\delta}^{\rm M} \ll 1}$, and thus ${C_{\text{tot}} \approx C_{\rm M}}$.
This suggests that the membrane primarily governs the capacitive response of the system, with only minor contributions from the diffuse layers.
Furthermore, at steady-state, the dimensionless total charge in each diffuse layer is ${2\bar{V}\bar{C}_\text{tot}}$. Since nonlinearities originate in the diffuse layers, this motivates our choice of the perturbation parameter, ${\eta=2\bar{V}\bar{C}_\text{tot}}$.

Considering Eqs.~\eqref{eq:tM} and~\eqref{eq:capacitance}, we may rewrite the relaxation timescale as ${\bar{\tau}_\mathrm{C} = \left(2\bar{L} + \Gamma\bar{\delta}^{\rm M} \right) \bar{C}_{\rm tot}}$, which implies a dimensionless resistance of ${2\bar{L} + \Gamma\bar{\delta}^{\rm M}}$.
In dimensional units, this gives a total resistance of
\begin{equation}
    \label{eq:resistance}
    R_{\rm tot} = \frac{\tau_\mathrm{C}}{C_\text{tot}}= \frac{\lambda_{\rm D}^2}{\epsilon D} \left( 2L + \Gamma\delta^\mathrm{M} \right) \ ,
\end{equation}
which consists of two contributions: the electrolyte resistance in both domains and an effective membrane resistance.

The idea of a membrane resistance may appear counterintuitive, since our model assumes the membrane to be impermeable to ions.
However, the effective membrane resistance arises from the continuity of the displacement field at the membrane-electrolyte interface. In fact, including such a resistance in the equivalent circuit is necessary to capture the initial jump in the membrane potential at ${t=0}$ (see lower right inset in Fig.~\ref{fig:phi-rho}(b)).
One may estimate this effective resistance as follows. The bulk electrolyte possesses a true conductivity due to ion motion, given by ${G=\epsilon D/\lambda_{\rm D}^2}$. An applied current $I$ then generates an electric field of magnitude ${I/G}$ in the electrolyte bulk. Due to the continuity of the displacement field, there must be an electric field in the membrane of magnitude ${\Gamma I/G}$, where $\Gamma$ is the dielectric mismatch. To represent this membrane field within a circuit model, one must include an apparent conductivity of ${G/\Gamma}$ for the membrane, despite the membrane itself being non-conducting. These two contributions, one from the bulk electrolyte and the other from the membrane, together recover the total resistance given in Eq.~\eqref{eq:resistance}. Notably, if the separating membrane is sufficiently thick or highly insulating, the dominant contribution to the resistance will come from the membrane.

In summary, Fig.~\ref{fig:equiv-circuit} shows the equivalent circuit corresponding to the linear relaxation regime. Each electrolyte domain contains two diffuse charge layers, each with capacitance $C_{\rm D}$, and a bulk domain with resistance ${R_{\rm L}=L/G}$. Since the length scales are well separated, ${\lambda_{\rm D}\ll L}$, we can separately consider the potential drop over each diffuse layer and the bulk electrolyte. The potential drop in the membrane arises from the combined contributions of a capacitance $C_{\rm M}$ and an effective resistance ${R_{\rm M}=\Gamma\delta^{\rm M}/G}$. The time-dependent response of this circuit recovers the membrane potential result given in Eq.~\eqref{eq:deltaV-tau}.

\begin{figure*}
    \centering
    {
        \hfill
        \includegraphics[width=0.495\linewidth]{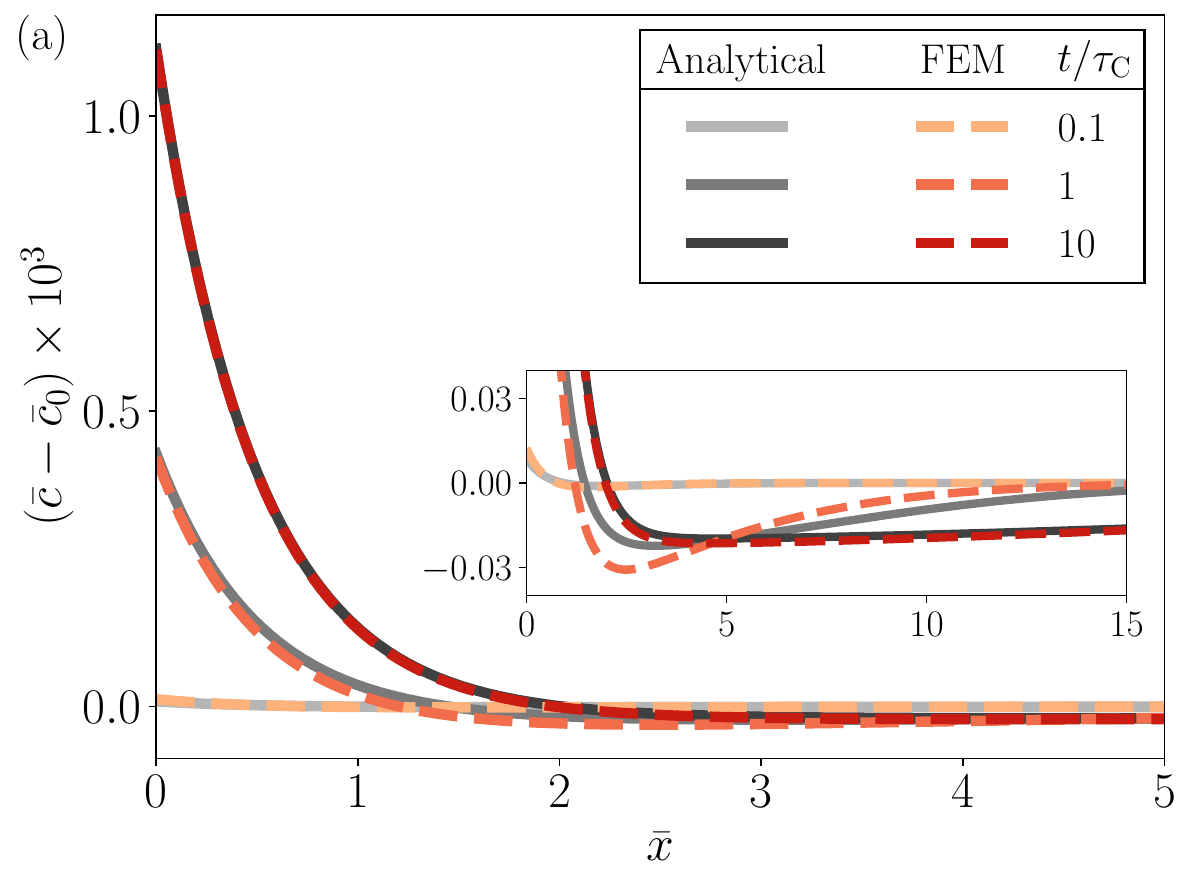}
        \hfill
        \includegraphics[width=0.495\linewidth]{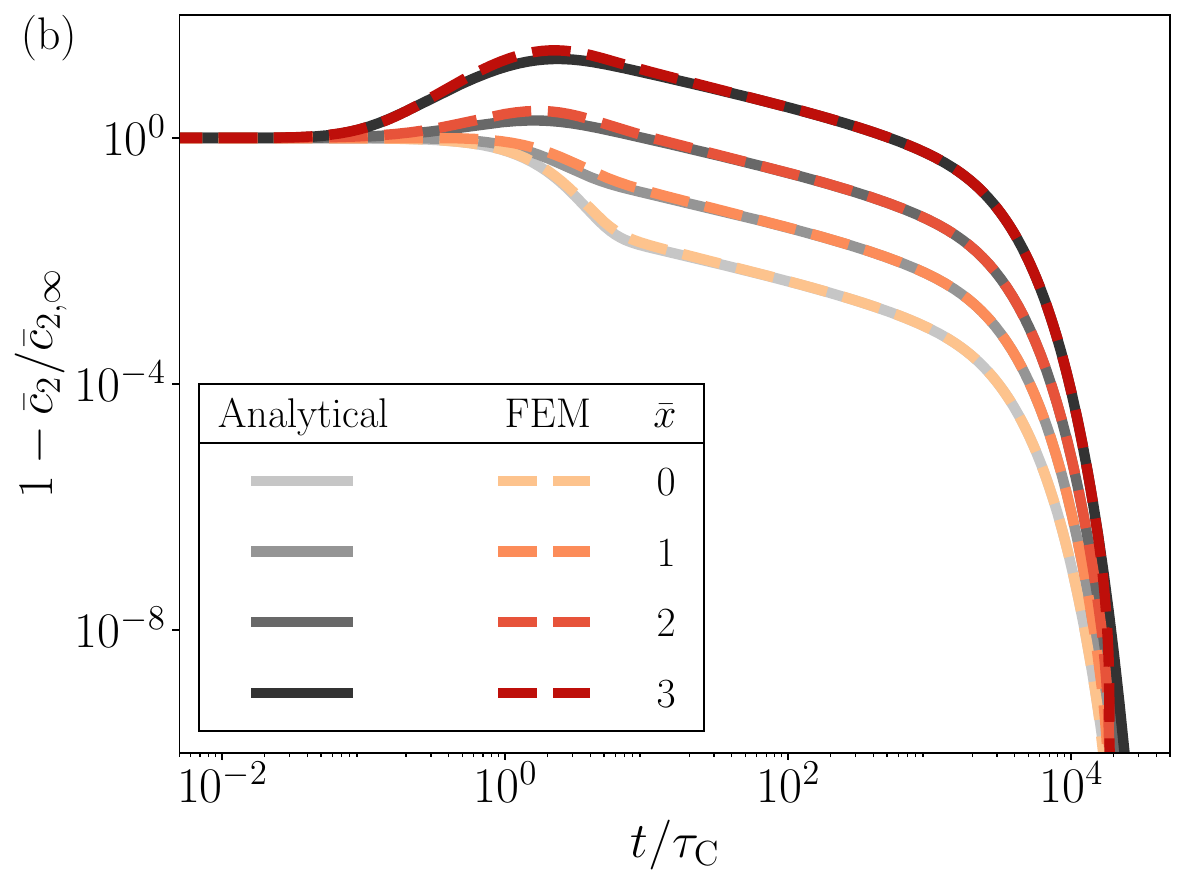}
        \hfill
    }
    \vspace{-0.5cm}
    \caption{
    (a) The deviation of the salt concentration, $\bar{c}$, from its bulk value, ${\bar{c}_0=1}$, shows accumulation in the diffuse charge layer compensated by depletion beyond the diffuse layer on the timescale $\tau_{\rm C}$. The inset highlights the depletion effect. (b)~The depletion converges to a uniform value across the entire bulk domain, with relaxation occurring on the diffusive timescale ${L^2/D}$. Parameters as in Fig.~\ref{fig:phi-rho}. The analytical solution plotted is the second-order contribution, $\eta^2\bar{c}_2$, with $\bar{c}_{2,\infty}$ being the steady-state value given by Eq.~\eqref{eq:c2-steady}.}
    \label{fig:conc-profiles}
\end{figure*}

\subsection{Nonlinear Dynamics}
We now proceed to analyze the nonlinear response to understand the emergent secondary relaxation process in Fig.~\ref{fig:transmembrane}. This requires analyzing the governing equations and boundary conditions up to second- and third-order in $\eta$.

$\\$
\textbf{Second Order:}
By the symmetry arguments, only the salt concentration has a nonzero contribution to second-order in $\eta$. Substituting the perturbation expansions Eqs.~\eqref{eq:perturb} in the governing equations, and collecting the $\mathcal{O}\left(\eta^2\right)$ terms results in the following diffusion equation for the salt concentration: 
\begin{subequations}\label{eq:second-order-eqns}
    \begin{equation}
        \label{eq:c2}
        \frac{\partial \bar{c}_2}{\partial \bar{t}} = \frac{\partial}{\partial \bar{x}}\left( \frac{\partial \bar{c}_2}{\partial \bar{x}} + 
 \bar{\rho}_1 \frac{\partial \bar{\phi}_1}{\partial \bar{x}}\right)\ ,
    \end{equation}
    subject to the boundary conditions
    \begin{equation}
        \label{eq:BC-c2}
        \left[ \frac{\partial \bar{c}_2}{\partial \bar{x}}  + \bar{\rho}_1 \, \frac{\partial \bar{\phi}_1}{\partial \bar{x}} \right]_{\bar{x}=0,\bar{L}} = 0\ ,
    \end{equation}
    and the initial condition ${\bar{c}_2 \left(\bar{x},0\right)=0}$.
\end{subequations}
According to Eq.~\eqref{eq:c2}, the concentration flux occurs to accommodate the charge accumulation in the diffuse charge layers.

Given the first-order solutions for $\bar{\rho}_1$ and $\bar{\phi}_1$, Eqs.~\eqref{eq:second-order-eqns}, in the limits of ${\bar{L} \gg \Gamma\bar{\delta}^{\rm M}}$ and ${\bar{t}\gg 1}$, can be recast as a Sturm-Liouville problem.
This yields the following solution (see Appendix~\ref{app:secondorder}),
\begin{align}
    \label{eq:c2-t}
    \begin{split}
        \bar{c}_2\left(\bar{x},\bar{t}\right) &= \frac{1}{2}\left( 1 - e^{-\bar{t}/\bar{\tau}_\mathrm{C}} \right)^2 \Bigg[  \left[f(\bar{x})\right]^2 \\
        &\quad\, - \frac{1}{\bar{L}} \sum_{n=-\infty}^\infty \frac{\cos \left(\alpha_n \bar{x}\right)}{1+\left(\alpha_n/2\right)^2} \exp \left( -\alpha_n^2 \bar{t} \right) \Bigg] \ ,
    \end{split}
\end{align}
where ${\alpha_n = 2n\pi/\bar{L}}$. 
Equation~\eqref{eq:c2-t} shows that the initial dynamics of the salt concentration evolves on the timescale $\tau_{\rm C}$, driven by the accumulation of charges in the diffuse layer, encapsulated by $f(\bar{x})$. This leads to an increase in salt concentration at all surfaces on the timescale $\tau_{\rm C}$, which is compensated by the depletion of salt in the neighborhood of the diffuse charge layer, as shown in Fig.~\ref{fig:conc-profiles}(a) and inset. The local depletion leads to salt concentration gradients, which relax diffusively within the bulk electrolyte until the concentration becomes uniform across the bulk domain. This occurs on the bulk diffusion timescale ${\tau_{\rm L}=\tau_{\rm D}/\alpha_1^2= L^2/\left(4\pi^2 D\right)}$, as seen from the leading-order contribution of the infinite series in Eq.~\eqref{eq:c2-t} and as shown in Fig.~\ref{fig:conc-profiles}(b). Figures~\ref{fig:conc-profiles}(a) and (b) compare the analytical solution with the numerical results,
showing qualitatively correct behavior at all times and quantitative agreement when ${t\gg \tau_{\rm C}}$.
The slight disagreement for ${t\lesssim \tau_{\rm C}}$ arises from neglecting nonhomogeneous terms, which become significant on timescales comparable to $\tau_{\rm C}$.\footnote{Using Eq.~\eqref{eq:phi1-t}, one can observe that the first-order electric field has two nonhomogeneous contributions, one from the diffuse layers and another from the linear bulk field which relaxes to zero.
We neglect the linear contribution, since it scales as ${1/\bar{\tau}_{\rm C}}$ and we assume ${\bar{\tau}_{\rm C}\gg 1}$ due to ${\bar{L}\gg\Gamma\bar{\delta}^{\rm M}}$. 
We also neglect an associated nonhomogeneous term that captures changes on the timescale $\bar{\tau}_{\rm C}$.
Due to these simplifications, the analytical solution is inexact when ${\bar{t}\lesssim \bar{\tau}_{\rm C}}$.
Details can be found in Appendix \ref{app:secondorder}.} Finally, the long time (${t\gg \tau_{\rm L}}$) behavior for the salt concentration saturates to
\begin{equation}
    \label{eq:c2-steady} \bar{c}_2\left(\bar{x},\bar{t}\to\infty\right) =\frac{1}{2} \left(e^{-2\bar{x}} + e^{-2\left(\bar{L}-\bar{x}\right)} - \frac{1}{\bar{L}}\right) \ ,
\end{equation}
which describes the accumulation of salt in each diffuse layer layer, compensated by uniform depletion throughout the electrolyte.

$\\$
\textbf{Third Order:}
We now analyze the third-order perturbation expansions in $\eta$, which yield the following evolution equations for $\bar{\rho}_3$ and $\bar{\phi}_3$: 
\begin{subequations}
\label{eq:third-order-eqs}
    \begin{align}
        \label{eq:rho3}
        \frac{\partial \bar{\rho}_3}{\partial \bar{t}} &= \frac{\partial^2 \bar{\rho}_3}{\partial \bar{x}^2} - \bar{\rho}_3 + \frac{\partial}{\partial \bar{x}} \left( \bar{c}_2 \frac{\partial \bar{\phi}_1}{\partial \bar{x}}\right) \ , \\
        \label{eq:phi3}
        \frac{\partial^2 \bar{\phi}_3}{\partial \bar{x}^2} &= - \bar{\rho}_3 \ ,
    \end{align}
\end{subequations}
subjected to boundary conditions
\begin{subequations}
    \begin{align}
        \label{eq:BC-rho3}
        \left[ \frac{\partial \bar{\rho}_3}{\partial \bar{x}} + \frac{\partial \bar{\phi}_3}{\partial \bar{x}} + \bar{c}_2 \, \frac{\partial \bar{\phi}_1}{\partial \bar{x}} \right]_{x=0,\bar{L}} &= 0 \ , \\
        \label{eq:BC-Robin3}
        \left.\frac{\partial \bar{\phi}_3}{\partial \bar{x}} \right|_{\bar{x}=0} &= \frac{2}{\Gamma\bar{\delta}^\mathrm{M}} \, \bar{\phi}_3 \left(0,\bar{t}\right) \ , \\
        \label{eq:BC-phi3}
       \bar{\phi}_3 \left(\bar{L},\bar{t}\right) &= 0 \ .
    \end{align}
\end{subequations}
From Eq.~\eqref{eq:third-order-eqs}, it can be seen that $\bar{\rho}_3$ and $\bar{\phi}_3$ are driven by the second-order salt concentration, $\bar{c}_2$, and the first-order electric potential, $\bar{\phi}_1$.

For ${t\lesssim \tau_{\rm C}}$, we expect the relaxation of the charge density and electric potential to be dominated by the first order term. Accordingly, in the analysis of the third-order solution, we consider only changes on the order ${t\sim \tau_{\rm L}}$, corresponding to the late-time variations in concentration.
On these timescales, any variations over $\tau_{\rm C}$ may be neglected, and so we remove all terms that scale as ${\exp \left(-\bar{t} / \bar{\tau}_{\rm C}\right)}$ from both $\bar{\phi}_1$ and $\bar{c}_2$ in Eqs.~\eqref{eq:phi1-t} and \eqref{eq:c2-t}, respectively.
Substituting these simplified forms into Eq.~\eqref{eq:rho3} yields
\begin{multline}
    \label{eq:rho3-h}
    \frac{\partial \bar{\rho}_3}{\partial \bar{t}} = \frac{\partial^2 \bar{\rho}_3}{\partial \bar{x}^2} - \bar{\rho}_3 - \frac{3}{2} \left[f\left(\bar{x}\right)\right]^3 \\
    + \frac{\partial}{\partial \bar{x}} \left[ \frac{f'\left(\bar{x}\right)}{2\bar{L}} \sum_{n=-\infty}^\infty \frac{\cos \left(\alpha_n \bar{x}\right)}{1+\left(\alpha_n/2\right)^2} \exp \left( -\alpha_n^2 \bar{t} \right) \right] \ ,
\end{multline}
where we use the identity ${f''\left(\bar{x}\right)=f\left(\bar{x}\right)}$ and the approximation  ${\left[f'\left(\bar{x}\right)\right]^2\approx\left[f\left(\bar{x}\right)\right]^2}$ when ${\bar{L}\gg 1}$.

Furthermore, when considering ${\bar{t}\gg\bar{\tau}_{\rm C}\gg 1}$, we recognize that only modes satisfying ${\alpha_n^2\ll 1}$ will contribute in Eq.~\eqref{eq:rho3-h}.
Therefore, the length scales of the cosine terms for these modes, ${1/\alpha_n}$, are much greater than 1. Since ${f\left(\bar{x}\right)}$ is non-negligible only on an ${\mathcal{O}\left(1\right)}$ region near ${\bar{x}=0}$ and ${\bar{x}=\bar{L}}$ and ${\cos\left(\alpha_n \bar{x}\right)=1}$ at both these points, we may approximate ${f'\left(\bar{x}\right)\cos\left(\alpha_n\bar{x}\right)\approx f'\left(\bar{x}\right)}$ for ${\bar{t}\gg \bar{\tau}_{\rm C}}$. Therefore, $\bar{x}$ dependence in the summation in Eq.~\eqref{eq:rho3-h} can be neglected, and the summation term reduces to its boundary value at ${\bar{x}=0,\bar{L}}$ given by
\begin{equation}
        \label{eq:YN}
        Y\left(\bar{t}\right) = \frac{1}{\bar{L}} \sum_{n=-\infty}^\infty \frac{\exp \left( -\alpha_n^2 \bar{t} \right)}{1+\left(\alpha_n/2\right)^2} \ .
\end{equation}
With these considerations, and again using the identity ${f''\left(\bar{x}\right)=f\left(\bar{x}\right)}$, Eq.~\eqref{eq:rho3-h} simplifies to
\begin{equation}
    \label{eq:rho3-h-simp}
    \frac{\partial \bar{\rho}_3}{\partial \bar{t}} = \frac{\partial^2 \bar{\rho}_3}{\partial \bar{x}^2} - \bar{\rho}_3 - \frac{1}{2} \left[ 3\left[f\left(\bar{x}\right)\right]^3 - f\left(\bar{x}\right) Y\left(\bar{t}\right) \right] \ ,
\end{equation}
and its boundary condition Eq.~\eqref{eq:BC-rho3} further simplifies to
\begin{equation}
    \label{eq:BC-rho3-h}
    \left[ \frac{\partial \bar{\rho}_3}{\partial \bar{x}} + \frac{\partial \bar{\phi}_3}{\partial \bar{x}} \right]_{\bar{x}=0,\bar{L}} = \frac{1}{2} \left[ Y\left(\bar{t}\right) - 1 \right] \ .
\end{equation}

Further simplifications can be made when  considering timescales ${\bar{t}\sim\bar{\tau}_{\rm L}}$. Recognizing that the time variation of the charge density ${\partial / \partial \bar{t}}$ varies as ${1/\bar{L}^2}$
in Eq.~\eqref{eq:rho3-h-simp}, we may further neglect the explicit time dependence of $\bar{\rho}_3$ and solve the following quasi-steady problem:
\begin{equation}
\label{eq:qs-rho3}
    0 = \frac{\partial^2 \bar{\rho}_3}{\partial \bar{x}^2} - \bar{\rho}_3 - \frac{1}{2} \left[ 3\left[f\left(\bar{x}\right)\right]^3 - f\left(\bar{x}\right) Y\left(\bar{t}\right) \right] \ .
\end{equation}
However, the quasi-steady problem is underdetermined, since $\bar{\rho}_3$ is determined by two Neumann boundary conditions in Eq.~\eqref{eq:BC-rho3-h}. This can be alleviated by closing the problem with the electroneutrality constraint
\begin{equation}
\label{eq:electroneutrality}
    \int_0^{\bar{L}}\bar{\rho}_3\left(\bar{x},\bar{t}\right) \, d\bar{x} = 0 \ ,
\end{equation}
for all times $\bar{t}$, which simply indicates that charge is conserved.

In summary, the governing equations for $\bar{\rho}_3$ and $\bar{\phi}_3$ of the reduced problem are given by Eqs.~\eqref{eq:qs-rho3} and \eqref{eq:phi3}, respectively, subjected to the boundary conditions Eqs.~\eqref{eq:BC-Robin3}, \eqref{eq:BC-phi3} and \eqref{eq:BC-rho3-h}, and the constraint Eq.~\eqref{eq:electroneutrality}. The exact solution to this system of equations is derived in Appendix~\ref{app:thirdorder}, and is given as
\begin{subequations}
\label{eq:third-order-soln}
    \begin{align}
            \label{eq:rho3-t}
            \begin{split}
            \bar{\rho}_3 \left(\bar{x},\bar{t}\right) &= \frac{1}{48} \left[ 9 \left[f\left(\bar{x}\right)\right]^3 - \left(\tfrac{4+3\Gamma\bar{\delta}^{\rm M}}{4+\Gamma\bar{\delta}^{\rm M}}\right) f\left(\bar{x}\right) \right] \\
            &\quad\, + \frac{1}{4} \left[ g\left(\bar{x}\right)- \left(\tfrac{8+\Gamma\bar{\delta}^{\rm M}}{4+\Gamma\bar{\delta}^{\rm M}}\right) f\left(\bar{x}\right) \right] Y\left(\bar{t}\right) \ ,
            \end{split}\\
            \label{eq:phi3-t}
            \begin{split}
            \bar{\phi}_3 \left(\bar{x},\bar{t}\right) &= \frac{1}{48} \left[ \left(\tfrac{4+3\Gamma\bar{\delta}^{\rm M}}{4+\Gamma\bar{\delta}^{\rm M}} \right)f\left(\bar{x}\right) - \left[f\left(\bar{x}\right)\right]^3 + \tfrac{2\Gamma\bar{\delta}^{\rm M}}{4+\Gamma\bar{\delta}^{\rm M}} \right] \\
            &\quad\, - \frac{1}{4} \left[ g\left(\bar{x}\right) + \left(\tfrac{\Gamma\bar{\delta}^{\rm M}}{4+\Gamma\bar{\delta}^{\rm M}}\right) f\left(\bar{x}\right) + \tfrac{\Gamma\bar{\delta}^{\rm M}}{4+\Gamma\bar{\delta}^{\rm M}} \right] Y\left(\bar{t}\right) \ ,
            \end{split}
        \end{align}
    where
    \begin{equation}
        \label{eq:gx}
        g\left(\bar{x}\right) = \bar{x}e^{-\bar{x}} - \left(\bar{L}-\bar{x}\right)e^{-\left(\bar{L}-\bar{x}\right)} \ .
    \end{equation}
\end{subequations}
We can interpret the third-order solutions as corrections to the linear boundary layer solution, $\bar{\rho}_1$ and $\bar{\phi}_1$, arising from interactions with the salt concentration. These corrections are necessary, since the diffuse layer follows a simple exponential profile only in the linear regime. Nonlinear corrections modify its structure, leading to a sharper diffuse layer, as evidenced by the ${\exp\left(-3\bar{x}\right)}$ term in ${\left[f\left(\bar{x}\right)\right]^3}$.
The time dependence of the charge density and electric potential arises from ${Y\left(\bar{t}\right)}$, which describes a diffusive relaxation process on a timescale of ${\tau_{\rm L}}$ and originates from the slow reorganization of the salt concentration over the bulk electrolyte.

Finally, the transmembrane potential in Eq.~\eqref{eq:def-transmembrane}, up to third order, can be obtained as 
\begin{equation}
    \label{eq:deltav-nonlinear}
    \bar{V}^\mathrm{M}_\text{non} \left(\bar{t}\right) = \bar{V}^{\rm M}_{\rm lin}(\bar{t}) + 
    \frac{\left(2\bar{V}\bar{C}_\text{tot}\right)^3}{1+4/\Gamma \bar{\delta}^\mathrm{M}} \left[ \frac{1}{6} - Y\left(\bar{t}\right) \right] \ ,
\end{equation}
where ${\bar{V}^{\rm M}_{\rm lin}}$ can be found from Eq.~\eqref{eq:deltaV-tau} as ${\bar{V}^{\rm M}_{\rm lin}=V^{\rm M}_{\rm lin}/\phi_{\rm T}}$.
As expected, the nonlinear correction is third-order in the applied voltage and in the limit of ${\Gamma\bar{\delta}^{\rm M}\gg 1}$, scales as ${\left(\Gamma\bar{\delta}^{\rm M}\right)^{-3}}$.

Figures~\ref{fig:phi-rho} and \ref{fig:transmembrane} show good agreement between the numerical results and the analytical solution, which includes both the first- and third-order terms. Notably, only the first-order terms contribute appreciably when plotting the charging dynamics directly as in Fig.~\ref{fig:phi-rho}. However, the contributions from both orders become evident when considering the transmembrane potential shown in Fig.~\ref{fig:transmembrane}. 
The first-order contribution, ${V_{\rm lin}^{\rm M}\left(t\right)}$, gives the initial relaxation behavior on timescales of $\tau_{\rm C}$, while the third-order contribution in ${V_{\rm non}^{\rm M}\left(t\right)}$ gives the long-time relaxation behavior on timescales of $\tau_{\rm L}$. Despite the approximations made to obtain the leading-order nonlinear correction, Fig.~\ref{fig:transmembrane} demonstrates the efficacy of the analytical solution over a wide range of parameters.

\subsection{Equilibrium Solution}
One may verify the consistency of the approximations in obtaining the third-order solutions by comparing them to the equilibrium behavior predicted by the Gouy-Chapman solution for electrical double layers~\cite{Gouy1910,chapman1913li}. To that end, in the limit of ${L\gg\lambda_{\rm D}}$, the electrolyte domain appears semi-infinite from each interface. Accordingly, we may approximate the nonlinear solution by the superposition of the Gouy-Chapman solution in each double layer. Due to the separation of length scales, these solutions interact only weakly as the double layers remain non-overlapping.

The Gouy-Chapman solution for an electrical double layer is given by
\begin{equation}
\label{eq:gc-potential}
    \bar{\phi}\left(\bar{x}\right)-\bar{\phi}_{\rm bulk} = 4\tanh^{-1}\left[\tanh\left(\frac{\bar{V}^{\rm D}_\infty}{4}\right) e^{-\bar{x}}\right] \ ,
\end{equation}
where $\bar{x}$ is the distance from the interface,  ${\bar{\phi}_{\rm bulk}}$ is the value of the potential as ${\bar{x}\to\infty}$, and $\bar{V}^{\rm D}_\infty$ is the potential difference across the diffuse layer. By symmetry, each of the four diffuse layers have the same structure and potential drop.
Using Eq.~\eqref{eq:gc-potential}, the electric field at the membrane-electrolyte interface is given by
\begin{equation}
    -\left.\frac{\partial \bar{\phi}}{\partial \bar{x}}\right|_{\bar{x}=0} = 2\sinh\left(\frac{\bar{V}_\infty^{\rm D}}{2}\right) \ ,
\end{equation}
which, by continuity of the dielectric displacement field, gives the equilibrium transmembrane potential:
\begin{equation}
\label{eq:memb-edl}
    \bar{V}_\infty^{\rm M} = 2\Gamma\bar{\delta}^{\rm M}\sinh\left(\frac{\bar{V}_\infty^{\rm D}}{2}\right) \ .
\end{equation}
The total applied potential difference is then
\begin{equation}
    2\bar{V} = \bar{V}_\infty^{\rm M} + 4\bar{V}_\infty^{\rm D} = \bar{V}_\infty^{\rm M} + 8\sinh^{-1}\left(\frac{\bar{V}_\infty^{\rm M}}{2\Gamma\bar{\delta}^{\rm M}}\right) \ .
\end{equation}
Expanding this expression about ${\bar{V}^{\rm M}_\infty=0}$ and using B\"{u}rmann's theorem \cite{whittaker1996course}, we can invert $\bar{V}^{\rm M}_\infty$ as a series in $\bar{V}$. This gives
\begin{equation}
\label{eq:eq-deltaV-exp}
    \bar{V}_\infty^{\rm M} = \Gamma\bar{\delta}^{\rm M}\eta+\frac{\Gamma\bar{\delta}^{\rm M}}{6\left(4+\Gamma\bar{\delta}^{\rm M}\right)}\eta^3 + \mathcal{O}\left(\eta^5\right) \ ,
\end{equation}
where ${\eta=2\bar{V}/\left(4+\Gamma\bar{\delta}^{\rm M}\right)}$ as defined in the perturbation expansion. We observe that the first term corresponds to $\bar{V}^{\rm M}_{\rm lin}$ in Eq.~\eqref{eq:deltaV-tau}, while the second corresponds to the second term of $\bar{V}^{\rm M}_{\rm non}$ in Eq.~\eqref{eq:deltav-nonlinear}, both in the limit ${\bar{t}\to\infty}$. Note that the expression in Eq.~\eqref{eq:eq-deltaV-exp} lacks the length scale correction in $\bar{V}^{\rm M}_{\rm non}$, since the Gouy-Chapman solution assumes ${\bar{L}\to\infty}$.

Lastly, we can substitute the expression for $\bar{V}_\infty^{\rm M}$ in Eq.~\eqref{eq:eq-deltaV-exp} into Eq.~\eqref{eq:memb-edl}, and Taylor-series expand it to find $\bar{V}_\infty^{\rm D}$ in powers of $\eta$. This can then be used in Eq.~\eqref{eq:gc-potential} to obtain the Gouy-Chapman potential in powers of $\eta$. The resulting expression, while not shown, agrees exactly with the steady-state portions of $\bar{\phi}_1$ and $\bar{\phi}_3$ in Eqs.~\eqref{eq:phi1-t} and \eqref{eq:phi3-t}, respectively,  except for the ${1/\bar{L}}$ correction arising from ${Y\left(\bar{t}\right)}$. Similarly, we may expand the Gouy-Chapman solution for the charge density ${\bar{\rho}=-\sinh\left(\bar{\phi}-\bar{\phi}_{\rm bulk}\right)}$ in $\eta$, which again matches the steady-state profiles of $\bar{\rho}_1$ and $\bar{\rho}_3$ as ${\bar{L}\to \infty}$. This agreement between the nonlinear steady-state solutions and the Gouy-Chapman solutions indicates that the analytical solution we obtain successfully captures the true charging dynamics leading to a steady state.

\section{Discussion}
\label{sec:discussion}

We investigate the timescales governing the charging dynamics of a model biological membrane-electrolyte system subjected to a step voltage. The applied potential difference leads to the formation of diffuse charge layers at the membrane and electrode interfaces, and steadily increases the transmembrane potential until equilibrium is reached. Through a perturbation analysis of the Poisson-Nernst-Planck equations, we show that the leading-order relaxation dynamics of this system is governed by the capacitive timescale $\tau_{\rm C}$ (Eq.~\eqref{eq:dim-tauC}), which grows linearly with the system size $L$. Following the initial regime, further relaxation of the system proceeds nonlinearly over the diffusion timescale ${\tau_{\rm L} \sim L^2/D}$, driven by the coupling of the diffuse charge layers to the reorganization of salt concentration throughout the bulk electrolyte. 

Importantly, because ${\tau_{\rm L} \gg \tau_{\rm C}}$, diffusion-mediated nonlinear effects always influence the approach to equilibrium, even when the applied voltage is small (${V\ll \phi_{\rm T}}$), as seen in Fig.~\ref{fig:transmembrane}. We estimate the onset of the nonlinear regime as 
\begin{equation}
    \label{eq:t_star}
    t^* \approx \tau_{\rm C} \ \ln \left[ \frac{\bar{L} \, \left( 2 + \Gamma\bar{\delta}^{\rm M}/2 \right)^3}{\bar{V}^2} \right] \ ,
\end{equation}
indicating that nonlinear effects becomes important for ${t\gg \tau_{\rm C}}$.
Since ${\tau_{\rm C}\sim L}$, nonlinear effects may be particularly important in confined geometries such as dendritic spines or vesicles, with length scales ${L\lesssim 100\text{ nm}}$ \cite{holcman2015new}. In these scenarios, diffusion effects and the ensuing nonlinearities may predominate within a microsecond.

Notably, the low capacitance of biological membranes substantially reduces $\tau_{\rm C}$ compared to $\tau_{\rm B}$, the relaxation timescale of a bare electrolyte \cite{Bazant04}. Under typical physiological conditions (${\lambda_{\rm D}\ll \Gamma\delta^{\rm M}\ll L}$), we find that ${\tau_{\rm C}\approx \left(\lambda_{\rm D}L/D\right) 2\lambda_{\rm D}/\Gamma\delta^{\rm M}}$, showing that in the presence of a biological membrane, the system charges and equilibriates faster by the factor ${\Gamma\delta^{\rm M}/\lambda_{\rm D}\sim 100}$.

We now highlight two representative biological examples highlighting why this rapid capacitive timescale plays an important role in membrane-charging phenomena. In their seminal voltage-clamp experiments on the squid giant axonal membrane, Hodgkin, Huxley and Katz extensively characterized millisecond-scale transmembrane currents associated with membrane excitability \cite{hodgkin1952measurement,hodgkin1952currents,hodgkin1952components,hodgkin1952dual,hodgkin1952quantitative}. However, for each voltage clamp setting, the authors documented an ``instantaneous surge of capacity current'' immediately after the membrane potential was changed. These capacity currents rapidly decayed to zero with a time constant of about ${6\text{ \textmu s}}$ \cite{hodgkin1952measurement}, and were followed by the slower, millisecond-scale currents associated with membrane excitability. In the simplest approximation, Hodgkin, Huxley, and Katz's experimental setup can be modeled as a lipid membrane placed between two electrodes, albeit with a cylindrical membrane geometry and a nonuniform separation distance (${L\sim 0.3 \text{ mm}}$ inside the axon and ${L\sim 2 \text{ mm}}$ outside). Using Eq.~\eqref{eq:dim-tauC} and taking 
${2L=2.3 \text{ mm}}$, ${D=1 \text{ nm}^2/\text{ns}}$, ${\lambda_{\rm D}=0.4 \text{ nm}}$, ${\Gamma = 20}$, and ${\delta^{\rm M}=5 \text{ nm}}$ yields ${\tau_{\rm C}\approx 3 \text{ \textmu s}}$, consistent with the value of ${6\text{ \textmu s}}$ reported by the authors. Furthermore, our circuit model (Eq.~\eqref{eq:resistance}) predicts an initial current of ${I= 2V/R_{\rm tot}}$; for an applied voltage of $40 \text{ mV}$, this gives ${I\approx 8 \text{ mA}/\text{cm}^2}$, aligning well with their measured peak current of ${I\approx 4.5 \text{ mA}/\text{cm}^2}$ \cite{hodgkin1952measurement}.\footnote{When fitting an exponential function to the data, the authors obtained a prefactor of $6.8 \text{ mA}/\text{cm}^2$, even closer to our theoretical estimate.} Crucially, the microsecond capacitive timescale ensured that the millisecond-scale transmembrane currents associated with membrane excitability were experimentally accessible, and not confounded by the capacitive currents. By contrast, the bare electrolyte timescale for their experimental setup (${\lambda_{\rm D}L/D\sim 0.2\text{ ms}}$) is within the timescale of membrane excitability. 

As another example, our recent work suggests that the capacitive timescale may govern membrane charging and discharging in a wider range of scenarios than captured in Fig.~\ref{fig:fig_schematic}. In Ref.~\cite{row2025spatiotemporal}, we analyzed the 3D electrochemical relaxation of a spatially localized transmembrane ionic current through a single ion transporter on a flat membrane. For a continuous current, we found that the transmembrane potential ${V^{\rm M}\left(r,t\right)}$ at a radial distance ${r\gg\Gamma\delta^{\rm M}}$ from the source relaxes according to
\begin{equation}
\frac{V^{\rm M}\left(r,t\right)}{V^{\rm M}_{\rm ss}\left(r\right)} = 1 -\left[1+\left(\frac{t}{\tau_{\rm C}^*\left(r\right)}\right)^2 \right]^{-\frac{1}{2}}  \ ,
\end{equation}
where ${V^{\rm M}_{\rm ss}\left(r\right)}$ is the steady-state limit of ${V^{\rm M}\left(r,t\right)}$, and ${\tau_{\rm C}^*\left(r\right)\approx \left(\lambda_{\rm D}r/D\right) 2\lambda_{\rm D}/\Gamma\delta^{\rm M}}$. This local version of $\tau_{\rm C}$ can be obtained from Eq.~\eqref{eq:dim-tauC} by replacing $L$ with $r$. Thus, even for a single, localized membrane current, a rapid capacitive timescale governs the local build-up of transmembrane potential. This result highlights the broader relevance of $\tau_{\rm C}$ for ion transport in biological membranes, governing how quickly regions around channels or transporters reach quasi-steady-state potentials. Understanding how these insights generalize to electrochemical relaxation in more complex geometries, such as spherical vesicles or cylindrical axons, and when nonlinearities manifest remains an important avenue for future work. 

$\\$\textbf{Authors' note:} We have become aware of recent independent work by Zhao \textit{et al.}~\cite{zhao2025diffuse} analyzing the charging dynamics of a capacitive membrane subjected to a voltage difference. While the authors of Ref.~\cite{zhao2025diffuse} considered electrolyte solutions of different ionic strengths separated by the membrane and despite the differences in the linear analysis, the resulting capacitive timescale agrees with the result presented in our article for the case of equal ionic strengths, albeit in the limit ${L \gg \Gamma \delta^{\rm M}}$. The analysis presented in this work extends beyond the linear regime and is applicable to applied voltages (${\sim 2.5\text{ V}}$), much higher than the thermal voltage (${\sim 25\text{ mV}}$), providing insights into the nonlinear regime and the timescales over which these effects manifest, which is relevant for many biological processes.

\section*{Acknowledgments}

The authors are grateful to Dr. Hyeongjoo Row and Dr. Sirui Ning for useful discussions.
J.B.F. acknowledges support from the U.S. Department of Energy, Office of Science, Office of Advanced Scientific Computing Research, Department of Energy Computational Science Graduate Fellowship under Award Number DE-SC0023112. K.K.M and J.B.F. are supported by Director, Office of Science, Office of Basic Energy Sciences, of the U.S. Department of Energy under contract No. DEAC02-05CH11231. K.S. and J.F. acknowledge support from the Hellman Foundation, the McKnight Foundation, and the University of California, Berkeley.

\appendix
\section{First-Order Solution}\label{app:firstorder}

Here, we derive the first-order analytical solution from Eqs.~\eqref{eq:rho1}--\eqref{eq:BC-phi1}.
We apply the separation of variables method and identify a primary relaxation timescale, $\tau_{\rm C}$, which we show to be significantly larger than all others, each of which is shorter than ${\tau}_{\rm D}$. Given this large separation, we provide the leading-order solution for $\bar{\rho}_1$ and $\bar{\phi}_1$ corresponding to the principal relaxation timescale.

Let ${\bar{\rho}_1^{\rm ss} \left(\bar{x}\right)}$ and ${\bar{\phi}_1^{\rm ss} \left(\bar{x}\right)}$ denote the first-order steady-state solutions obtained by setting ${\partial/\partial \bar{t} = 0}$ in Eqs.~\eqref{eq:first-order}, subjected to the boundary conditions Eqs.~\eqref{eq:first-order-BCs}. In the limit ${\bar{L}\gg 1}$, making use of the identity ${f''\left(\bar{x}\right)=f\left(\bar{x}\right)}$, one can obtain ${\bar{\rho}_1^{\rm ss} \left(\bar{x}\right) = f\left(\bar{x}\right)}$ and ${\bar{\phi}_1^{\rm ss} \left(\bar{x}\right) = \Gamma\bar{\delta}^{\rm M}/2 + 1 - f\left(\bar{x}\right)}$, where ${f\left(\bar{x}\right)}$ is given by Eq.~\eqref{eq:f}.

Subtracting ${\bar{\rho}_1^{\rm ss}}$ from the first-order solution of the charge density and expressing the difference using separation of variables:
\begin{equation}
    \label{eq:SL-rho1}
    \bar{\rho}_1 \left(\bar{x}, \bar{t}\right) - \bar{\rho}_1^{\rm ss} \left(\bar{x}\right) = X\left(\bar{x}\right) T\left(\bar{t}\right)\ ,
\end{equation}
and substituting into Eq.~\eqref{eq:rho1}, we obtain,
\begin{equation}
    \label{eq:SL}
    \frac{T'}{T} = \frac{X''}{X} - 1 = -\mu^2\ .
\end{equation}
Here, we only consider positive values, $\mu^2$, based on the physical assumption that the solution may not diverge as ${\bar{t}\to\infty}$. Accordingly, using Eq.~\eqref{eq:SL}, the electric potential can be expressed generally as
\begin{equation}
    \label{eq:SL-phi1}
    \bar{\phi}_1 \left(\bar{x}, \bar{t}\right) - \bar{\phi}_1^{\rm ss} \left(\bar{x}\right) = \left[\frac{X\left(\bar{x}\right)}{\mu^2 - 1} + E\left(\bar{t}\right) \bar{x} + F\left(\bar{t}\right)\right] T\left(\bar{t}\right)\ ,
\end{equation}
obtained by integrating Eq.~\eqref{eq:phi1}.
Here, ${E\left(\bar{t}\right)}$ and ${F\left(\bar{t}\right)}$ are arbitrary functions of time that arise as integration constants---we will find from the boundary conditions that they must be constant in time.

Applying the boundary conditions from Eqs.~\eqref{eq:first-order-BCs} gives
\begin{align}
    \label{eq:BC-X1}
    \frac{\mu^2}{\mu^2-1}X'\left(0\right) + E\left(\bar{t}\right) &= 0 \ , \\
    \label{eq:BC-X2}
    \frac{\mu^2}{\mu^2-1}X'\left(\bar{L}\right) + E\left(\bar{t}\right) &= 0 \ , \\
    \label{eq:BC-X3}
    \frac{X'\left(0\right)}{\mu^2-1} + E\left(\bar{t}\right) &= \frac{2}{\Gamma\bar{\delta}^{\rm M}}\left[\frac{X\left(0\right)}{\mu^2-1} + F\left(\bar{t}\right)\right] \ , \\
    \label{eq:BC-X4}
    \frac{X\left(\bar{L}\right)}{\mu^2-1} + E\left(\bar{t}\right)\bar{L} + F\left(\bar{t}\right) &= 0 \ ,
\end{align}
from which we can immediately recognize that ${E\left(\bar{t}\right)}$ and ${F\left(\bar{t}\right)}$ must be constant. By eliminating ${E\left(\bar{t}\right)}$ and ${F\left(\bar{t}\right)}$ we find
\begin{align}
    X'\left(0\right) &= X'\left(\bar{L}\right) \ , \\
    X\left(0\right) - X\left(\bar{L}\right) &= X'\left(0\right) \left[\frac{\Gamma\bar{\delta}^{\rm M}}{2}\left(1-\mu^2\right)-\bar{L}\mu^2\right] \ ,
\end{align}
as boundary conditions on ${X\left(\bar{x}\right)}$. 

Equation~\eqref{eq:SL} gives an exponentially decaying solution in time
\begin{equation}
    \label{eq:SL-T}
    T \left(\bar{t}\right) = \exp \left( -\mu^2 \bar{t} \right)\ .
\end{equation}
In space, however, the structure of the solution depends on whether ${0<\mu<1}$, ${\mu=1}$, or ${\mu > 1}$.

First, we consider the case of ${0<\mu<1}$. We then find
\begin{equation}
    \label{eq:SL-X1}
    X \left(\bar{x}\right) = A_0 \sinh \left( \sqrt{1 - \mu^2} \ \bar{x}\right) + B_0 \cosh \left( \sqrt{1 - \mu^2} \ \bar{x}\right)\ ,
\end{equation}
where $A_0$ and $B_0$ are undetermined constants.
Applying the boundary conditions on $X$ gives the transcendental equation for $\mu$ as
\begin{equation}
    \label{eq:SL-transcendental-1}
    \bar{L} \mu^2 - \frac{\Gamma\bar{\delta}^{\rm M}}{2} \left(1 - \mu^2\right) = \frac{2}{\sqrt{1 - \mu^2}} \tanh \left( \frac{\bar{L}}{2} \sqrt{1 - \mu^2} \right)\ ,
\end{equation}
which, when $\bar{L}$ is large (i.e., ${e^{-\bar{L}} \approx 0}$) and ${\mu \ll 1}$ can be approximated by
\begin{equation}
    \label{eq:SL-transcendental-simp}
    \bar{L} \mu^2 - \frac{\Gamma\bar{\delta}^{\rm M}}{2} \left(1 - \mu^2\right) = 2 + \mu^2\ .
\end{equation}
This has only one solution, given by
\begin{equation}
    \label{eq:mu0}
    \mu_0^2 = \frac{4 + \Gamma\bar{\delta}^{\rm M}}{2\bar{L} - 2 + \Gamma\bar{\delta}^{\rm M}}\ ,
\end{equation}
which is much smaller than 1 for ${\bar{L}\gg \Gamma\bar{\delta}^{\rm M}}$. Under our assumption ${\bar{L}\gg 1}$, we may neglect the 2 in the denominator. In this limit, it is immediately evident that

\begin{equation}
    \frac{1}{\mu_0^2} = \frac{2\bar{L} + \Gamma \bar{\delta}^{\rm M}}{4 + \Gamma \bar{\delta}^{\rm M}} = \bar{\tau}_{\rm C}
\end{equation}

Next, if ${\mu=1}$, we find
\begin{equation}
\label{eq:SL-x-lin}
    X\left(\bar{x}\right) = A_1\bar{x}+B_1 \ .
\end{equation}
Note that in this case, $\bar{\phi}_1$ takes a different functional form, since the one presented in Eq.~\eqref{eq:SL-phi1} assumes ${\mu\neq 1}$. This also leads to different boundary conditions for $X$, which are not satisfied by Eq.~\eqref{eq:SL-x-lin} for general $\bar{L}$ and $\Gamma\bar{\delta}^{\rm M}$. Therefore, ${\mu=1}$ is not an eigenvalue.

Now, we consider the case of ${\mu>1}$. Here, the solution in space can be written as
\begin{equation}
    \label{eq:SL-X2}
    X\left(\bar{x}\right) = A_2 \sin \left( \sqrt{\mu^2 - 1} \ \bar{x}\right) + B_2 \cos \left( \sqrt{\mu^2 - 1} \ \bar{x}\right)\ .
\end{equation}
Applying the boundary conditions on $X$ gives the transcendental equation for $\mu$ as
\begin{equation}
    \label{eq:SL-transcendental-2}
    \bar{L} \mu^2 + \frac{\Gamma\bar{\delta}^{\rm M}}{2} \left(\mu^2 - 1\right) = \frac{2}{\sqrt{\mu^2 - 1}} \tan \left( \frac{\bar{L}}{2} \sqrt{\mu^2 - 1} \right)\ .
\end{equation}
Note that this equation has infinite solutions, and for ${\bar{L}\gg1}$, they approach the singular points of the tangent term. Therefore, by taking the approximation ${\bar{L} \, \sqrt{\mu^2 - 1} / 2 \approx n\pi + \pi/2}$ we find the eigenvalues as
\begin{equation}
    \label{eq:mun}
    \mu_n^2 \approx 1 + \left[ \frac{\left(2n+1\right)\pi}{\bar{L}} \right]^2\ ,
\end{equation}
where ${n \in \{1, 2, 3, \dots\}}$.

Given the set of eigenvalues ${\mu_0,\mu_1, \mu_2,\ldots}$, one may write a series solution in terms of the eigenfunctions associated with Eqs.~\eqref{eq:SL-X1} and \eqref{eq:SL-X2}, obtaining the complete analytical solution. 
However, since ${\mu_0\ll\mu_1<\ldots}$, we note that the leading-order relaxation timescale is given by ${\bar{\tau}_{\rm C} = 1 / \mu_0^2}$. Furthermore, for ${n\geq 1}$, we have ${\mu_n>1}$, so the associated relaxation times are all on the order of the Debye time or shorter.
Therefore, the relaxation behavior at times ${\bar{t}\gg 1}$ is dominated by the leading-order term corresponding to ${\bar{\tau}_{\rm C}\gg 1}$ as ${\mu_0\ll 1}$. In this case, the asymptotic solution for ${t\gg \tau_{\rm D}}$ can be written as
\begin{equation}
    \label{eq:SL-rho1-ans}
    \bar{\rho}_1 \left(\bar{x},\bar{t}\right) - \bar{\rho}_1^{\rm ss} \left(\bar{x}\right) \sim \left[ e^{-\sqrt{1-\mu_0^2}\left(\bar{L}-\bar{x}\right)} - e^{-\sqrt{1-\mu_0^2}\,\bar{x}} \right] e^{-\mu_0^2 \bar{t}} \ ,
\end{equation}
which can be further approximated as
\begin{equation}
    \label{eq:SL-rho1-ans-simp}
    \bar{\rho}_1 \left(\bar{x},\bar{t}\right) - \bar{\rho}_1^{\rm ss} \left(\bar{x}\right) \sim - \left[ e^{-\bar{x}} - e^{-\left(\bar{L}-\bar{x}\right)} \right] e^{-\bar{t}/\bar{\tau}_{\rm C}}\ ,
\end{equation}
using ${\mu_0 \ll 1}$. Considering the initial condition ${\bar{t}\to 0}$, given as ${\bar{\rho}_1\left(\bar{x},0\right)=0}$, the constant of proportionality in Eq.~\eqref{eq:SL-rho1-ans-simp} is 1, yielding the solution presented in Eq.~\eqref{eq:rho1-t}. 

Using the boundary conditions in Eqs.~\eqref{eq:BC-X1}--\eqref{eq:BC-X4}, we find ${E\left(\bar{t}\right)}$ and ${F\left(\bar{t}\right)}$ to be
\begin{align}
    E\left(\bar{t}\right) &= \mu_0^2 \ ,\\
    F\left(\bar{t}\right) &= 1 - \mu_0^2 \bar{L} \ ,
\end{align}
in the limits ${\mu_0\ll 1}$ and ${\bar{L}\gg 1}$.
Substituting these expressions into Eq.~\eqref{eq:SL-phi1} gives the solution for $\bar{\phi}_1$, as presented in Eq.~\eqref{eq:phi1-t}.

\section{Second-Order Solution}
\label{app:secondorder}

In this section, we derive the second-order solution for the salt concentration governed by  Eq.~\eqref{eq:second-order-eqns}. Substituting the leading-order solutions for $\bar{\rho}_1$ and $\bar{\phi}_1$ yields the following governing equation:
\begin{align}
    \label{eq:c2-g}
    \begin{split}
        \frac{\partial \bar{c}_2}{\partial \bar{t}} = \frac{\partial^2 \bar{c}_2}{\partial \bar{x}^2} &- 2 \left[f\left(\bar{x}\right)\right]^2 \left( 1 - e^{-\bar{t}/\bar{\tau}_{\rm C}} \right)^2 \\
        &+ f'\left(\bar{x}\right) \frac{e^{-\bar{t}/\bar{\tau}_{\rm C}}}{\bar{\tau}_{\rm C}} \left( 1 - e^{-\bar{t}/\bar{\tau}_{\rm C}} \right) \ ,
    \end{split}
\end{align}
where we approximate ${[f'(\bar{x})]^2\approx [f(\bar{x})]^2}$ under the assumption ${e^{-\bar{L}}\approx  0}$. The boundary conditions in Eq.~\eqref{eq:BC-c2} give 
\begin{equation}
    \label{eq:BC-c2-g}
    \frac{\partial \bar{c}_2}{\partial \bar{x}} \bigg|_{\bar{x}=0,\bar{L}} \pm \left( 1 - e^{-\bar{t}/\bar{\tau}_{\rm C}} \right)^2 \pm \frac{e^{-\bar{t}/\bar{\tau}_{\rm C}}}{\bar{\tau}_{\rm C}} \left( 1 - e^{-\bar{t}/\bar{\tau}_{\rm C}} \right) = 0 \ ,
\end{equation}
where the positive sign corresponds to ${\bar{x} = 0}$, and the negative sign corresponds to ${\bar{x} = \bar{L}}$.

The problem as stated in Eqs.~\eqref{eq:c2-g} and \eqref{eq:BC-c2-g} is a nonhomogeneous heat equation.
The solution to this problem can be expressed exactly as a convolution in space and time with the nonhomogeneous term. However, to facilitate interpretability, we simplify the problem by considering the limits of ${\bar{\tau}_{\rm C}\gg 1}$ and ${\bar{t}\gtrsim \bar{\tau}_{\rm C}}$. This gives a simplified solution that provides physical insight into the timescales corresponding to the secondary relaxation process.

Under the assumption ${\bar{\tau}_{\rm C} \gg 1}$, i.e., ${\bar{L} \gg \Gamma\bar{\delta}^{\rm M}}$, the problem simplifies to
\begin{equation}
    \label{eq:c2-g-simp}
    \frac{\partial \bar{c}_2}{\partial \bar{t}} = \frac{\partial^2 \bar{c}_2}{\partial \bar{x}^2} - 2 \left[f\left(\bar{x}\right)\right]^2 \left( 1 - e^{-\bar{t}/\bar{\tau}_{\rm C}} \right)^2 \ ,
\end{equation}
subjected to the boundary conditions
\begin{equation}
    \label{eq:BC-c2-g-simp}
    \frac{\partial \bar{c}_2}{\partial \bar{x}} \bigg|_{\bar{x}=0,\bar{L}} \pm \left( 1 - e^{-\bar{t}/\bar{\tau}_{\rm C}} \right)^2 = 0 \ ,
\end{equation}
with initial condition ${\bar{c}_2(\bar{x},0)=0}$.
The form of Eqs.~\eqref{eq:c2-g-simp} and~\eqref{eq:BC-c2-g-simp} motivates a change of variables to
\begin{equation}
    \label{eq:u}
    u\left(\bar{x}, \bar{t}\right) = \frac{2\bar{c}_2 \left(\bar{x}, \bar{t}\right)}{\left( 1 - e^{-\bar{t}/\bar{\tau}_{\rm C}} \right)^2} - \left[f\left(\bar{x}\right)\right]^2 \ ,
\end{equation}
which transforms the governing equation to
\begin{equation}
    \label{eq:u-governing}
    \frac{\partial u}{\partial \bar{t}} = \frac{\partial^2 u}{\partial \bar{x}^2} - \frac{2 \, e^{-\bar{t}/\bar{\tau}_{\rm C}}}{\bar{\tau}_{\rm C} \left(1 - e^{-\bar{t}/\bar{\tau}_{\rm C}} \right)} \left(u + \left[f\left(\bar{x}\right)\right]^2\right) \ ,
\end{equation}
where we again approximate ${[f'(\bar{x})]^2\approx [f(\bar{x})]^2}$. Using the assumption ${\bar{\tau}_{\rm C}\gg 1}$ and considering ${\bar{t} \gg 1}$, the nonhomogeneous term may be neglected. This yields a regular Sturm-Liouville problem
\begin{equation}
    \label{eq:u-governing-simp}
    \frac{\partial u}{\partial \bar{t}} = \frac{\partial^2 u}{\partial \bar{x}^2} \ ,
\end{equation}
subjected to the boundary conditions
\begin{equation}
    \label{eq:u-BC}
    \frac{\partial u}{\partial \bar{x}}\bigg|_{\bar{x} = 0,\bar{L}} = 0 \ ,
\end{equation}
and initial condition
\begin{equation}
    \label{eq:u-IC}
    u\left(\bar{x},0\right) = - \left[f\left(\bar{x}\right)\right]^2 \ .
\end{equation}
The solution to this Sturm-Liouville problem is
\begin{equation}
    \label{eq:u-t}
    u\left(\bar{x}, \bar{t}\right) = - \frac{1}{\bar{L}} \sum_{n=-\infty}^\infty \frac{\cos \left(\alpha_n \bar{x}\right)}{1+\left(\alpha_n\right/2)^2} \, \exp \left( -\alpha_n^2 \bar{t} \right) \ ,
\end{equation}
where ${\alpha_n = 2n\pi/\bar{L}}$.
Therefore, the second-order correction in concentration can be calculated by
\begin{equation}
    \label{eq:c2-u}
    \bar{c}_2\left(\bar{x}, \bar{t}\right) = \frac{1}{2} \left( \left[f\left(\bar{x}\right)\right]^2 + u\left(\bar{x}, \bar{t}\right) \right) \left( 1 - e^{-\bar{t}/\bar{\tau}_{\rm C}} \right)^2 \ ,
\end{equation}
as given in Eq.~\eqref{eq:c2-t}.

\section{Third-Order Solution}
\label{app:thirdorder}

We end with deriving the third-order solution for $\bar{\rho}_3$ and $\bar{\phi}_3$ governed by Eqs.~\eqref{eq:qs-rho3} and 
\eqref{eq:phi3}, boundary conditions \eqref{eq:BC-rho3-h}, \eqref{eq:BC-Robin3} and \eqref{eq:BC-phi3}, along with the constraint \eqref{eq:electroneutrality}.
Recognizing that the quasi-steady governing equation for $\bar{\rho}_3$, given by Eq.~\eqref{eq:qs-rho3}, is an ODE in $\bar{x}$, its solution takes the form
\begin{multline}
    \label{eq:rho3-const}
    \bar{\rho}_3\left(\bar{x},\bar{t}\right) = A\left(\bar{t}\right)f\left(\bar{x}\right) + B\left(\bar{t}\right)f'\left(\bar{x}\right) \\
    + \frac{3}{16}\left[f\left(\bar{x}\right)\right]^3 + \frac{1}{4}g\left(\bar{x}\right)Y\left(\bar{t}\right) \ .
\end{multline}
Here, ${A\left(\bar{t}\right)}$ and ${B\left(\bar{t}\right)}$ are arbitrary functions of time arising from the general solution and the final two terms satisfy the particular solution. The function $g\left(\bar{x}\right)$ is given by Eq.~\eqref{eq:gx} in the main text. Note, as before, we approximate ${\left[f'\left(\bar{x}\right)\right]^2\approx \left[f\left(\bar{x}\right)\right]^2}$ under the assumption ${e^{-\bar{L}} \approx 0}$. Applying the constraint in Eq.~\eqref{eq:electroneutrality} then gives ${B\left(\bar{t}\right)=0}$.

Substituting the expression for $\bar{\rho}_3$ from Eq.~\eqref{eq:rho3-const} into the governing equation for $\bar{\phi}_3$ (Eq.~\eqref{eq:phi3}), and solving the resulting ODE yields
\begin{multline}
    \label{eq:phi3-const}
    \bar{\phi}_3\left(\bar{x},\bar{t}\right) = C\left(\bar{t}\right)\bar{x} + D\left(\bar{t}\right)
    -A\left(\bar{t}\right)f\left(\bar{x}\right) \\
    - \frac{1}{48}\left[f\left(\bar{x}\right)\right]^3
    - \frac{1}{4}\left[g\left(\bar{x}\right)+2f\left(\bar{x}\right)\right]Y\left(\bar{t}\right) \ ,
\end{multline}
Here, ${C\left(\bar{t}\right)}$ and ${D\left(\bar{t}\right)}$ are arbitrary functions of time arising from the general solution with the remaining terms corresponding to the particular solution.

Now, we use the boundary conditions to resolve the three remaining constants, ${A\left(\bar{t}\right)}$, ${C\left(\bar{t}\right)}$, and ${D\left(\bar{t}\right)}$. Applying Eq.~\eqref{eq:BC-rho3-h} gives
\begin{align}
    \begin{split}
    \left[ \frac{\partial \bar{\rho}_3}{\partial \bar{x}} + \frac{\partial \bar{\phi}_3}{\partial \bar{x}} \right]_{\bar{x}=0,\bar{L}} &= \frac{1}{2}\left[Y\left(\bar{t}\right)-1\right] + C\left(\bar{t}\right) \\
    &= \frac{1}{2}\left[Y\left(\bar{t}\right)-1\right] \ ,
    \end{split}
\end{align}
which clearly implies ${C\left(\bar{t}\right)=0}$. Next, from the Dirichlet boundary condition on $\bar{\phi}_3$ at ${\bar{x}=\bar{L}}$, given by Eq.~\eqref{eq:BC-phi3}, we have

\begin{equation}
\label{eq:AD-eq1}
    D\left(\bar{t}\right) + A\left(\bar{t}\right) + \frac{1}{48}+\frac{1}{2}Y\left(\bar{t}\right) = 0 \ .
\end{equation}
Finally, applying the Robin boundary condition at ${\bar{x}=0}$, given by Eq.~\eqref{eq:BC-Robin3}, implies
\begin{multline}
\label{eq:AD-eq2}
    A\left(\bar{t}\right)+\frac{1}{16}+\frac{1}{4}Y\left(\bar{t}\right) \\
    = \frac{2}{\Gamma\bar{\delta}^{\rm M}}\left[D\left(\bar{t}\right) - A\left(\bar{t}\right) - \frac{1}{48}-\frac{1}{2}Y\left(\bar{t}\right)\right] \ .
\end{multline}
Equations \eqref{eq:AD-eq1} and \eqref{eq:AD-eq2} comprise a linear system in two variables for $A(\bar{t})$ and $D(\bar{t})$, to which the solution is given by
\begin{align}
    A\left(\bar{t}\right) &= -\frac{1}{48}\left(\frac{4+3\Gamma\bar{\delta}^{\rm M}}{4+\Gamma\bar{\delta}^{\rm M}}\right) - \frac{1}{4}\left(\frac{8+\Gamma\bar{\delta}^{\rm M}}{4+\Gamma\bar{\delta}^{\rm M}}\right)Y\left(\bar{t}\right) \ ,\\
    D\left(\bar{t}\right) &= \frac{1}{24}\left(\frac{\Gamma\bar{\delta}^{\rm M}}{4+\Gamma\bar{\delta}^{\rm M}}\right) - \frac{1}{4}\left(\frac{\Gamma\bar{\delta}^{\rm M}}{4+\Gamma\bar{\delta}^{\rm M}}\right)Y\left(\bar{t}\right)\ .
\end{align}
Substituting the above expressions into Eqs.~\eqref{eq:rho3-const} and \eqref{eq:phi3-const} and simplifying yields Eqs.~\eqref{eq:rho3-t} and \eqref{eq:phi3-t} in the main text, respectively.

\bibliography{refs}

\end{document}